\documentclass{article}
\pdfoutput=1

\usepackage{arxiv}

\usepackage{amsmath}
\DeclareMathOperator*{\argmax}{arg\,max}

\usepackage{algorithm}
\usepackage{algcompatible}
\usepackage{adjustbox}
\usepackage{amsfonts}
\usepackage{amssymb}
\usepackage{amsbsy}
\usepackage[T1]{fontenc}
\usepackage{textcomp}
\usepackage{graphicx}
\usepackage{listings}
\usepackage{multirow}
\usepackage[normalem]{ulem}
\usepackage{color}
\definecolor{gray}{rgb}{0.5,0.5,0.5}
\usepackage{enumitem}
\usepackage{graphicx}
\graphicspath{{./}}

\usepackage[hidelinks]{hyperref} 
\usepackage{fixltx2e}
\usepackage{soul}
\usepackage{tabularx}
\usepackage{longtable}
\usepackage{afterpage} 
\usepackage{colortbl}
\newcolumntype{Y}{>{\centering\arraybackslash}X}
\usepackage{tikz}

\newcommand{\ignore}[1]{}
\newcommand*{\circled}[1]{\textcircled{\footnotesize #1}}

\usepackage[utf8]{inputenc} 

\usepackage{booktabs}       
\usepackage{microtype}      

\title{Subsurface Boundary Geometry Modeling: Applying Computational Physics, Computer Vision and Signal Processing Techniques to Geoscience}
\shorttitle{Subsurface boundary geometry modeling from an interdisciplinary perspective}

\date{November 2, 2019}

\author{
The final version published in \textit{IEEE Access} is available at\\ \url{https://doi.org/10.1109/ACCESS.2019.2951605}\\ \\
  \textbf{Raymond~Leung}\vspace{2mm} \\
  Australian Centre for Field Robotics (ACFR)\\
  Faculty of Engineering\\
  The University of Sydney\\
  Sydney, NSW 2006 \\
  \texttt{raymond.leung@sydney.edu.au} \\
}

\renewcommand{\headeright}{Leung}

\begin{document}
\maketitle

\begin{abstract}
This paper describes an interdisciplinary approach to geometry modeling of geospatial boundaries. The objective is to extract surfaces from irregular spatial patterns using differential geometry and obtain coherent directional predictions along the boundary of extracted surfaces to enable more targeted sampling and exploration. Specific difficulties of the data include sparsity, incompleteness, causality and resolution disparity. Surface slopes are estimated using only sparse samples from cross-sections within a geological domain with no other information at intermediate depths. From boundary detection to subsurface reconstruction, processes are automated in between. The key problems to be solved are boundary extraction, region correspondence and propagation of the boundaries via contour morphing. Established techniques from computational physics, computer vision and signal processing are used with appropriate modifications to address challenges in each area. To facilitate boundary extraction, an edge map synthesis procedure is presented. This works with connected component analysis, anisotropic diffusion and active contours to convert unordered points into regularized boundaries. For region correspondence, component relationships are handled via graphical decomposition. FFT-based spatial alignment strategies are used in region merging and splitting scenarios. Shape changes between aligned regions are described by contour metamorphosis. Specifically, local spatial deformation is modeled by PDE and computed using level-set methods. Directional predictions are obtained using particle trajectories by following the evolving boundary. However, when a branching point is encountered, particles may lose track of the wavefront. To overcome this, a curvelet backtracking algorithm has been proposed to recover information for boundary segments without particle coverage to minimize shape distortion.
\end{abstract}

\keywords{Interdisciplinary Perspective\and Active Contours\and Backtracking\and Contour Morphing\and Directional Prediction\and Particle Trajectories\and Spatial Correspondence\and Subsurface Boundaries\and Wavefront Propagation.
\newline\newline
\textbf{CCS Concepts}:\newline \hspace{8mm}$\bullet$ Computing methodologies\,$\rightarrow$\, Computer graphics\,$\rightarrow$\, Shape modeling\,$\rightarrow$\, \textbf{Parametric curve and surface models};\newline$\bullet$ Computing methodologies\,$\rightarrow$\ldots\,Computer vision representations\,$\rightarrow$\, \textbf{Shape representation};\newline$\bullet$ Computing methodologies\,$\rightarrow$\ldots\, Computer vision problems\,$\rightarrow$\, \textbf{Tracking};\newline$\bullet$ Applied computing\,$\rightarrow$\, Physical sciences\,$\rightarrow$\, \textbf{Earth sciences}.
}
\ignore{

}

\begin{table}
\centering
\begin{tabular}{c}
\includegraphics[width=75mm]{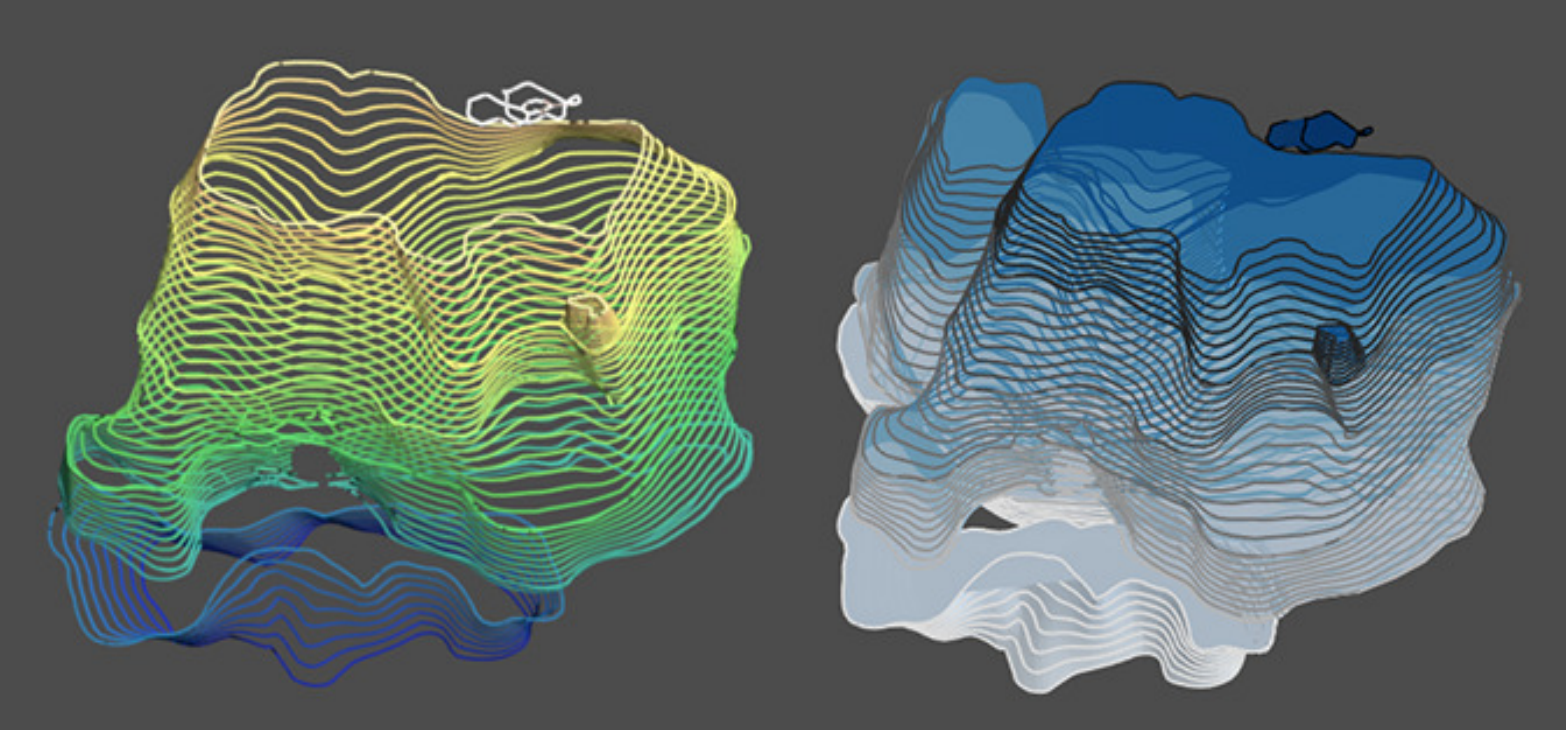}
\end{tabular}
\end{table}

\section{Introduction}\label{sect:bgm-introduction}
This paper considers the feasibility of modeling geospatial boundaries using differential geometry given sparse observations. The objective is to extract surfaces from spatial patterns and obtain coherent directional predictions along the boundary of the extracted surfaces. Modeling underground geological formations is challenging in general because the measurements are sparse and indirect. Due to operational constraints and the significant costs associated with data gathering\footnote{This includes operator cost, energy expenditure for drilling, replacement cost of mechanical parts, efficiency cost of coordinating dependent processes such as blasting and excavation, and the cost of performing chemical assays to determine the composition, material type or geological domain associated with each sample.}, the available observations may not paint a complete picture in terms of spatial coverage. These measurements, although point-based, differ from those encountered in computer vision or image processing in some signficant ways. The input consists of sparse spatial patterns in the form of irregularly spaced labeled drilled holes. Not only are the measurement locations sparse in the x-y plane, the sampling is less dense along the z-axis. Geo-assay data is typically collected sequentially on a bench-by-bench basis from the top down (each bench has a height of $\sim$10m) with significant time lapse in-between. In contrast, pixels are captured almost instantaneously using an image sensor array. The combination of these two factors means volumetric segmentation approaches that utilize 3D partial derivatives, and those that conduct minimal path search with respect to z, are not applicable as there are no voxels or fine-grain data available at intermediate depths between successive benches.

From a system perspective, the output provides a volumetric reconstruction of subterranean surfaces that conveys directional information. The motivation for predicting the slope of extracted boundaries is to provide guidance for more targeted drilling and exploration. The accuracy of this directional information needs only be commensurate with the resolution of the raw input (roughly $\sim$5m) for it to be useful in a mining context. However, the directional estimates need to be coherent. An example of what not to do is using the outward normals of a contour as a means for extrapolation which inevitably cross-over when non-convex boundaries are involved. Hence, partial differential equations (PDE) are used to describe boundary movement in a more principled manner.

The input data used in this work contrasts with dense data sources such as point cloud produced by terrain laser scanners (LIDAR) \cite{buckley-08}\cite{frank-07} and high-resolution slices generated by computed tomography \cite{nilsson-05} and presents its own challenges. Data incompleteness, sparsity, causality and resolution disparity are some of the issues to contend with and probable reasons for why differential geometry has not been more widely used in subterranean geology modeling. Most established PDE techniques for modeling surfaces operate on uniform grid data whereas our input samples are irregularly spaced. To overcome this, a bridging step based on boundary detection and edge map synthesis is described. This enables level-set methods to be applied to contours extracted from sparse non-uniform data points.

\begin{figure*}[!htb]
\centering
\includegraphics[width=165mm]{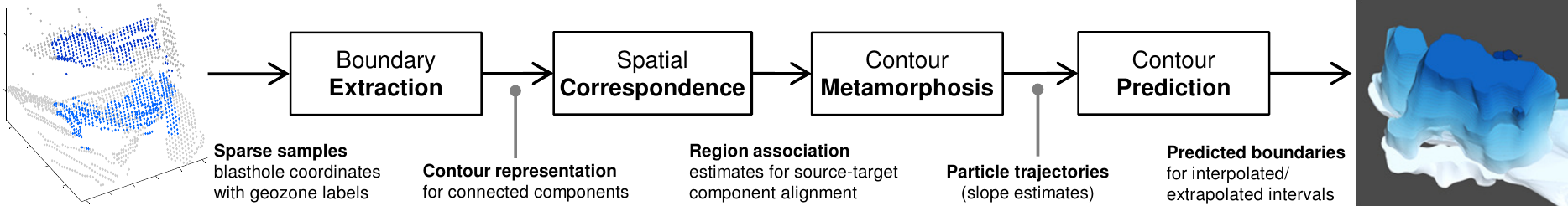}
\caption{Processing pipeline -- a high level abstraction of the Boundary Geometry Modeling (BGM) framework.}
\label{fig:pipeline}
\end{figure*}

As an overview, Fig.~\ref{fig:pipeline} shows the processes for achieving the ultimate goal. Once a set of contours emerge following boundary extraction, the work flow next enters the spatial correspondence (reasoning) phase whose purpose is to associate `source' regions with `target' regions at successive intervals and estimate component displacements. This problem shares many similarities with object tracking in computer vision, but is made difficult by significant variations rather than gradual changes in contour shape. Often, the vertical resolution is low, whilst some regions may not be sampled adequately, or at all, due to operational constraints. A region association and translation estimation approach is proposed to deal with the complexity of region merging and splitting from a resource allocation perspective.

Each set of associated source--target contours define a region in a motion-compensated frame for which contour morphing (spatial warping) \cite{nilsson-05} is applied. The objective is to model the shape of the boundary in between two cross-sections as a propagating interface. The underlying premise is that local surface deformation can be described as an evolutionary process governed by some PDE. Although the exact form used in this work might not match reality in terms of ore genesis, it is a reasonable alternative to unconstrained warping approaches and does provide a continuous deformable model. Using level-set methods \cite{adalsteinsson-95}, topological changes can be handled seamlessly during the morphing process. To facilitate slope prediction, the evolving interface (contour boundaries) are tracked using particle trajectories. This works well along portions of the boundaries where differences in curvature between the source and target contours are small. It breaks down when branching occurs, i.e., when a curvelet emanates from a single point. To address this, a novel backtracking algorithm has been proposed to recover lost information during particle advection and minimize boundary distortion attributed to tracking failure.

\subsection{Related works}
This paper is distinguished from prior work through its attempt in fully automating a chain of processes required to reconstruct subterranean surfaces using sparse labeled data. These processes include boundary extraction, spatial correspondence and contour evolution, as outlined in Fig.~\ref{fig:pipeline}. Although the contours for various geological domains (henceforth referred as \textit{geozones}) can be specified interactively, our model basically requires only a set of unordered points, sampled non-uniformly across an orebody in multiple cross-sections as input.

In \cite{sprague-05}, Sprague and de Kemp presented a partially automated tool which uses B{\'e}zier and NURBS (non-uniform rational B-spline) curves to model non-planar 3D surfaces. Fitting under the guidance of a control frame, it makes use of 2D \textit{interpreted} plan views provided by mine geologists, created using local slices of projected drill hole data at semi-regular spaced depths. The authors emphasized that in moving from 2D polygonal lines to 3D surface construction, the continuation of common features requires delicate spatial synchronization across neighboring plan-sections, and manual correspondence performed by the geologist was critical to the integrity of the modeled structure and preventing self-intersections. The term \textit{map trace} was defined as a ``geological interpretation delineating the intersection of a geologic boundary as it breaks through the surface\ldots or as it intersects a given elevation plane''. This definition closely resembles our notion of \textit{active contour extracted boundaries} which represent segmented regions in a geozone. Their technique were refined by imposing positional and orientation constraints using structural ribbons (expert knowledge), thus it may be categorized as a semi-supervised technique.

In \cite{mitavsova-93}, Mit{\'a}{\v{s}}ov{\'a} and Hofierka applied differential geometry, more specifically, regularized spline with tension to topographic analysis of a watershed. This utilized convex/concave sections with smooth curvature for interpolation. However, the focus was to model the top surface (as opposed to sub-surfaces) using dense elevation data obtained via remote sensing (as opposed to sparse data).

In \cite{kaufmann-08}, Kaufmann and Martin built 3D subsurface models using a variety of sources (drill holes, cross-sections and geological maps) with different motivations. Their goal was to further understand subsoil characteristics such as hydrogeologic or geothermic properties of the geological bodies. Their surfaces were modeled using DSI (discrete smooth interpolation) which computes the location of nodes by balancing roughness and misfit constraints (see Mallet \cite{mallet-92} and Frank \cite{frank-07}). This is representative of a class of information-rich GIS fusion approaches which utilize topographic, geological and structural data \cite{zanchi-09}. In contrast, our approach makes the most of the limited data in an information-poor environment where only blasthole locations and geozone labels are available.

In \cite{caumon-09}, Caumon et al.\,presented general guidelines for creating a 3D structural model made of faults and horizons using sparse field data. Their focus was natural resource evaluation and hazard assessment; triangulated surfaces with variable resolution was also discussed. The authors offered many insights, one notable comment is that ``3D subsurface modeling is generally not an end, but a means of improving data interpretation through visualization\ldots to generate support for numerical simulations of complex phenomena (i.e., earthquakes, fluid transport) in which structures \cite{dewaele-18} play an important role.'' This is also true of subsurface models serving as decision support tools in mining and geological exploration.

In \cite{dirstein-13}, Dirstein et al.\,demonstrated that automated surface extraction and segmentation of peak and troughs from seismic survey can provide insight into structural and sedimentary morphology. In particular, differential geometry was used to select objects of concave and convex curvature, these features can help identify subtle cues for fluid flow events that are perhaps over-looked by conventional interpretation methods. For an in-depth survey of past efforts and current interests in 3D geological mapping, readers are referred to \cite{berg-09}, \cite{berg-11}, \cite{lindsay-13} and \cite{lin-17}. Relevant techniques from computational physics and computer vision are presented in Appendixes~A and B. As a quick overview, the major themes and related works are highlighted in Table~\ref{tab:computational-physics-computer-vision-themes}.

\begin{table}[h]
\centering
\caption{Computational physics and computer vision techniques used in this work (refer to Appendixes A--B in the \href{https://ieeexplore.ieee.org/ielx7/6287639/8600701/8891690/supplementalmaterial.pdf?tp=&arnumber=8891690}{\color{blue}Supplementary Material})}\label{tab:computational-physics-computer-vision-themes}
\begin{tabular}{|c|l|}\hline
& Description and referenced works\\ \hline
A & Theoretical treatment of active contours \& gradient vector field\\
 &\quad (Kass et al., 1988) \cite{kass-88}, (Caselles et al., 1997) \cite{caselles-97},\\
 &\quad (Ivins and Porrill, 1995) \cite{ivins-95}, (Xu and Prince, 1998) \cite{xu-98},\\
 &\quad (Horn and Schunck, 1981) \cite{horn-81}.\\ \hline
B & Foundations for contour metamorphosis\\
 &\quad (Nilsson et al., 2005) \cite{nilsson-05}, (Breen and Whitaker, 2001) \cite{breen-01},\\
 &\quad (Museth et al., 2005) \cite{museth-05}, (Nielsen and Museth, 2006) \cite{nielsen-06},\\
 &\quad (Bertalmio et al., 2000) \cite{bertalmio-00}, (Osher and Sethian, 1988) \cite{osher-88},\\
 &\quad (Peng et al., 1999) \cite{peng-99}, (Nagashima et al., 2007) \cite{nagashima-07}\\ \hline
\end{tabular}
\end{table}

In both Dirstein's and our work, curvature preserving spatial representations are used to good effect albeit the objectives and modalities are different. One aspect of Dirstein's work is waveform analysis, where genetic fitness (similarity) relative to a genotype (waveform signature) can reveal relative stability of neighboring segments above and below a particular surface and confer understanding about its structure and stratigraphy. In our work, we obtain an essentially continuous representation for the slope of geozone subsurfaces from sparse boundary patterns through contour extraction and metamorphosis; this provides useful directions for subsequent drilling and exploration. Our motivation stems from an interest in integrating techniques from pattern analysis, computational physics, computer vision and signal processing, bringing these to bear on a subsurface reconstruction problem, solving it in unconventional ways.

\subsection{Application Scenario}
The target application is the modeling of subsurface boundaries in an open-pit mine of sedimentary iron ore deposits.\footnote{The most common minerals are hematite $\text{Fe}^{3+}_2\text{O}_3$ and goethite  Fe\textsuperscript{3+}O(OH). These deposits are believed to have formed as chemical precipitates on the floor of shallow marine basins in a highly oxidizing environment during the Proterozoic eon, circa 1.9--2.4 billion years ago.} Here, the sparse spatial patterns derive from  blasthole samples where each is labeled with a geozone. The \textit{geozone} labels provide a classification of geological domains based on mineralization and stratigraphic units \cite{balamurali-16}. From a mining perspective, these are used to segregate high-grade minerals from low grade or unprofitable materials. With some effort, the geozones can be turned into coherent spatial clusters. However, as unorganized points, or even with a discretized block-based representation, there is no easy way of obtaining a coherent motion field, one that describes the direction a boundary moves in consistently without artifacts or discontinuities. This is one major reason for using a curve-based, deformable surface model.

At the outset, it is important to distinguish this work from 3D segmentation approaches \cite{ardon-06,ardon-07} that use energy functional for minimum path optimization given contours at specific depths which is a common scenario in computer tomography. These approaches often require 3D partial derivatives to be computed, our data simply do not have the requisite (vertical) resolution to facilitate that. Furthermore, as our data are literally mined through manual processes, they only become available progressively at periodic intervals. In our application, slope trends need to be estimated in a causal manner using cross-section samples from as little as two successive benches (at two particular depths). Our problem also differs from grade estimation or material classification for which a variety of probabilistic inference and machine learning approaches are known \cite{goodfellow-12}, \cite{tahmasebi-12}, \cite{crackell-14}, \cite{dirstein-13}, \cite{balamurali-16}, \cite{silversides-16}, \cite{karpatne-18}, \cite{jafrasteh-18}. The focus here is explicit spatial representation which encompasses boundary extraction, region correspondence and capturing changes to subsurfaces using differential geometry. 

\begin{figure}[!htb]
\centering
\includegraphics[width=85mm]{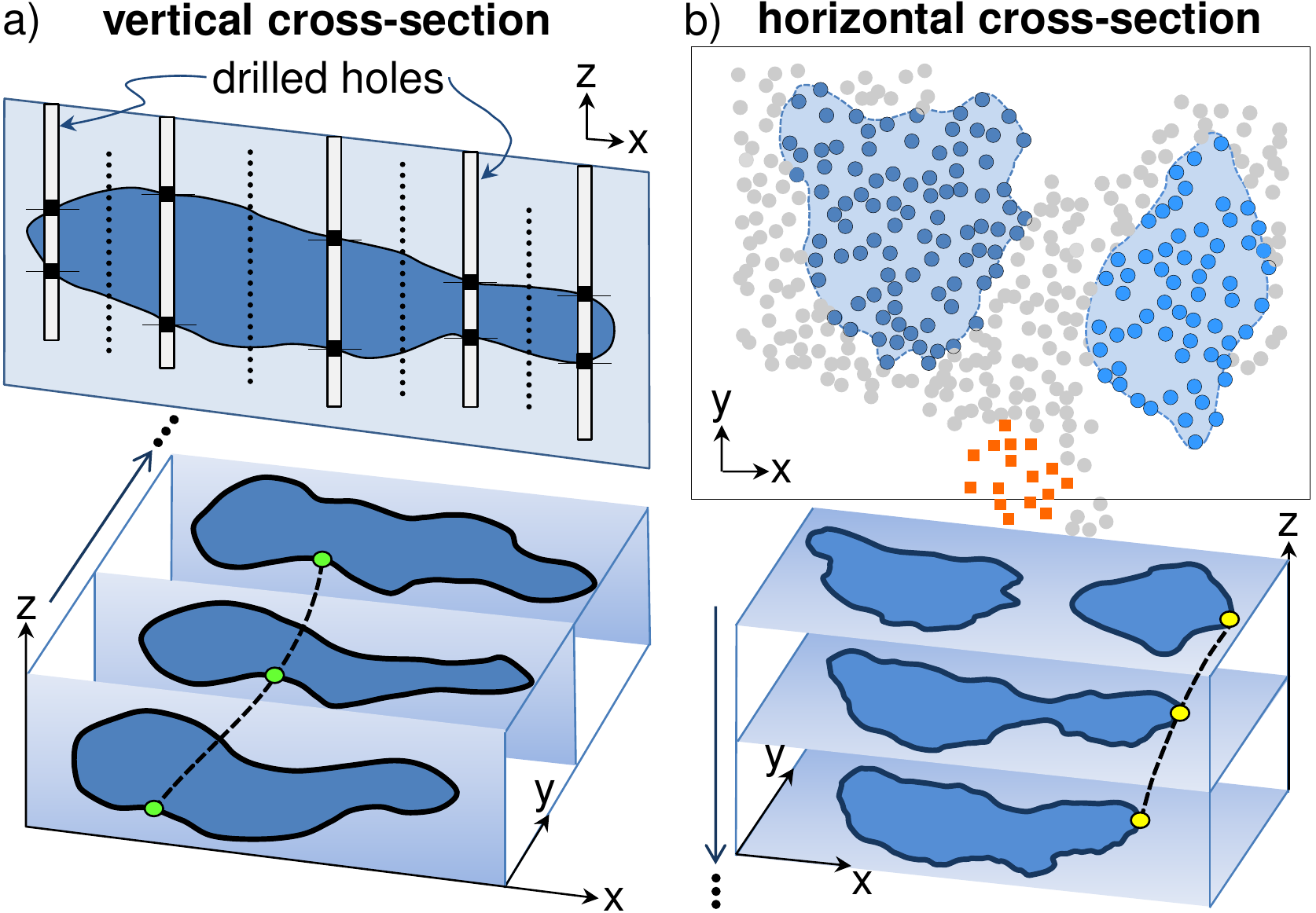}
\caption{Application scenarios. (a) In mining, holes are drilled perpendicular to the strike of the orebody. Geozone transitions (black squares) are commonly established by identifying marker bands using measure-while-drilling or assay data and/or visual inspection of rock specimen. Boundary is interpolated at locations (dotted lines) between the drilled holes. The estimated boundary is subsequently extrapolated and joined in the orthogonal direction, i.e., between successive vertical cross-sections to generate an approximate 3D orebody surface model. (b) Blastholes are more densely and non-uniformly sampled in the x-y plane. In this case, the boundaries shown in the horizontal cross-sections are not actually given. Instead, they are obtained using the proposed boundary extraction techniques. Modeling how the volume evolves through the cross-sections (as hinted by the dash lines) and estimating the slopes are the primary objectives of this work.}
\label{fig:drill-hole-contours}
\end{figure}

The proposed system takes input in one of two forms. The contours which describe geospatial regions are either given or derived. Fig.~\ref{fig:drill-hole-contours}(a) depicts a common mining scenario where stratigraphic information are obtained by examining core specimen extracted from the drilled holes. Alternatively, geozone transitions (marker bands) are established from geochemical assays or geophysical measurements taken while drilling. The main characteristics of this data is high resolution in z and sparse sampling in the x-y plane. Geologists typically interpolate the boundary at locations (dotted lines) between the drilled holes based on an understanding of the geological setting. The time and effort required to ``join up'' these vertical slices to form a preliminary 3D surface model can be substantial. Fig~\ref{fig:drill-hole-contours}(b) depicts a second scenario where horizontal cross-sections are taken. This scenario differs from the previous through denser sampling in the x-y plane and having relatively low z-resolution; blastholes are plentiful albeit non-uniformly sampled. In this case, the contours are \textit{not} given --- only the coordinates and geozone designations (classification labels) of the blastholes are known. This second scenario poses additional challenges with respect to \textit{boundary extraction} which will be further considered in this paper. These pictures highlight the main objective of this work which is to model how a volumetric region evolves through the cross-sections using contours and estimate the direction in which the boundary moves in.

The proposed boundary geometry modeling framework (henceforth, abbreviated as BGM) consists of four sub-systems. The processing pipeline is shown in Fig.~\ref{fig:pipeline}. The role of each subsystem is briefly described below.
\begin{itemize}
\item \textit{Boundary extraction}\ finds connected regions (components) from labeled blastholes. It locates boundary samples on each component and converts them into a closed contour. 
\item \textit{Spatial correspondence}\ automates the process of region association and component alignment. Having identified one or more components on each cross-section, the next goal is to match-up and motion-compensate contours, i.e., find the optimal translation which brings corresponding regions into alignment. By convention, contours found in the top and bottom of any two successive slices are referred as `source' and `target' components respectively.
\item \textit{Contour metamorphosis}\ uses particle trajectories to facilitate slope estimation. The goal is to model residual differences after source--target components are aligned in a common frame. Contour boundaries are embedded in a 2D level-set function, by tracking the movement of the zero-interface, one establishes pathways for morphing the source region(s) into the target region(s) as the level-set evolves under the dynamics of a PDE.
\item \textit{Contour prediction}\ reconstructs a 3D volume through interpolation or extrapolation of normalized trajectories.
\end{itemize}

\subsection{Contributions}
The main contributions are as follows.
\begin{itemize}
\item For boundary extraction, an edge map synthesis procedure is described which converts sparse non-uniform data points to an image representation. Gradient vector field and active contours are used to extract shapes (regularized contours) from unordered edge pixels.
\item For spatial correspondence, region association and component alignment problems are solved using FFT cross-correlation under spatial constraints. Obstacle avoidance is approached from a resource allocation view point, multiple-source multiple-target component relationships are explored using intersection graphs.
\item For contour metamorphosis, slope estimation is achieved using particle trajectories. A curvelet backtracking algorithm has been devised to overcome tracking failure caused by branching; a phenomenon that leads to boundary distortion. This occurs when there is a significant mismatch in shape (local curvature) between the corresponding boundary segments.
\end{itemize}

\begin{table}[!htb]
\footnotesize
\centering
\caption{Techniques employed in Boundary Geometry Modeling}
\label{tab:caterlogue-techniques}
\begin{tabular}{|l|l|}
\hline
Technique & Field of origin\,/\,inspiration\\ \hline
\multicolumn{2}{|c|}{\textsc{Boundary Extraction}}\\
Connected component analysis & data analysis, clustering\\
Orientation selective gap closure & mathematical morphology\\
Synthesis of edge structures & image processing\\
GVF (anisotropic diffusion) & computational physics\\
Active contours (object localization) & computer vision\\ \hline
\multicolumn{2}{|c|}{\textsc{Spatial Correspondence}}\\
Region association & probability\,/\,statistics\\
Cross-correlation/shift estimation & signal processing (FFT)\\
Component alignment strategies: & resource allocation,\\
\quad resolve conflicts, avoid obstacles & perception\\
Source--target dependency tree & graphical decomposition\\
Land mass characterization & shape analysis\\ \hline
\multicolumn{2}{|c|}{\textsc{Contour Metamorphosis\,/\,Prediction}}\\
Level-sets, signed distance function & applied mathematics\\ 
Wave propagation (hyperbolic PDE) & computational physics\\
Particle trajectories & object tracking\\
Boundary tracing & topology, computer vision\\
Surface reconstruction from slices & computer graphics\\ \hline
\end{tabular}
\end{table}

Table~\ref{tab:caterlogue-techniques} provides an overview of the techniques employed in each area. In the ensuing sections, the reasons for choosing these techniques will be elaborated as specific challenges are described. We discuss how standard approaches are modified to address particular needs. The solutions sought for different parts of the problem draw inspiration from different fields.

\section{Boundary extraction} \label{sect:bgm-boundary-extraction}
The first objective is to extract the boundary of contiguous spatial regions (connected components) given sparse data. For input, we have the spatial coordinates of blastholes for various horizontal cross-sections, and these blastholes have been classified as belonging to different geozones from prior analysis based on material properties or chemical composition. An example of this is shown in Fig.~\ref{fig:drill-hole-contours}(b) where the blue, gray dots and orange squares represent blastholes sampled from three different geozones. An important point about these samples is that they are irregularly spaced and unordered. Consequently, efficient algorithms for labeling connected components \cite{shapiro-stockman-02,suzuki-03} that perform sequential scans on a uniform grid cannot be used here. A related problem is that attempts at forming an edge map by connecting peripheral samples often produce false edges and loops which make clean boundary extraction difficult. These issues will be addressed in due course. In preparation for what follows, the boundary extraction process is summarized by a series of steps.
\begin{itemize}
\item \textbf{Connected component analysis}: For each cross-section, identify separate clusters within each geozone.
\item \textbf{Blasthole boundary detection}: Find samples located on the boundary of each cluster\,/\,connected component.
\item \textbf{Edge map synthesis}: Project the region (edges connecting the boundary samples) onto an image grid.
\item \textbf{Gradient vector field computation}: Drive the active contour (PDE solution) toward the boundary.
\item \textbf{Active contours evolution}: Extract the boundary as a closed contour described by $N$ control points.
\end{itemize}

\subsection{Connected component analysis} \label{sect:bgm-connected-component-analysis}
The modification involves using a kD-tree for nearest neighbor search and adopts a region growing approach. All samples from the same geozone and same cross-section are numbered from 1 to $n$, initially each belonging to the cluster of `self'. Neighboring samples within a radius of $r$ are merged with the current sample and labeled with the minimum cluster index amongst the group. Cluster membership information is propagated iteratively until no further changes occur and $S$ connected components remain.

\subsection{Blasthole boundary detection} \label{sect:bgm-blasthole-boundary-detection}
For each connected component, boundary samples are identified by thresholding the local entropy which is significantly non-zero at geozone transition points. Suppose a sample $n$ has $N_n$ neighbors within a radius of $r$ and the fraction of samples belonging to geozones $g_1$ and $g_2$ are $p_{n,1}$ and $p_{n,2}$. The local entropy is computed as $h_n = -\sum_i p_{n,i}\log_2 (p_{n,i}+\varepsilon)$. Sample $n$ is marked as a boundary sample if $h_n\ge \max\{T_\text{entropy}, h^\text{(median)}_n\}$ where $T_\text{entropy}=0.5$ and $h^\text{(median)}_n$ [the median entropy in $n$'s neighborhood] is used to suppress ``non-maximum'' responses.

\subsection{Automated gap closure} \label{sect:bgm-edge-completion}
\begin{figure}[!htb]
\centering
\includegraphics[width=136mm]{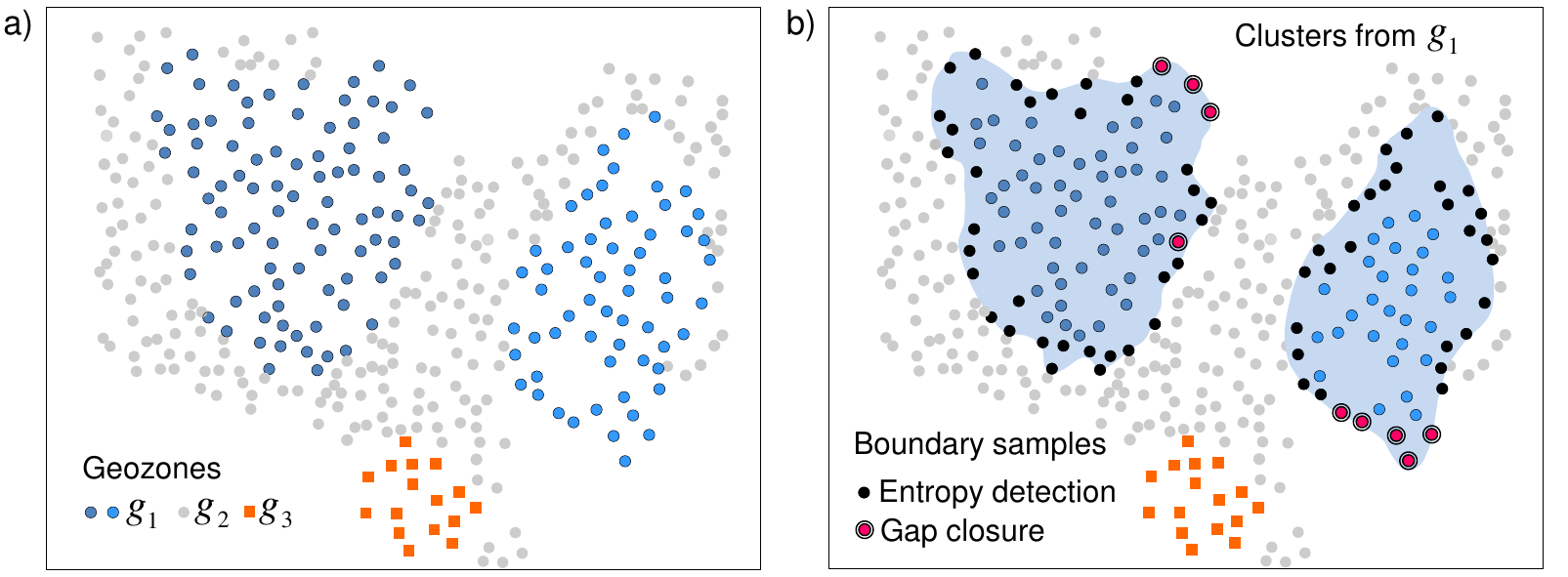}
\caption{Connected component analysis and blasthole boundary detection. (a) Input provides blasthole coordinates and geozone labels, one horizontal cross-section is shown. (b) Two clusters (connected components) have been identified in geozone $g_1$; black dots represent boundary samples detected by thresholding local entropy, red dots represent gap closure informed by orientation analysis.}
\label{fig:cca-boundary-detection}
\end{figure}

One limitation of this detector is the entropy tends to zero when the geozone labels no longer vary. This creates a problem as boundaries remain essentially open at the frontiers of surveyed regions which are not bordered by a different geozone. To remedy this situation, we perform orientation analysis to close these gaps. The objective is to recognize samples on the outskirts of a geozone as edges. The direction of the $K_\text{orient}$ closest neighbors from $n$ are computed and sorted in ascending order. If a gap larger than $T_\text{orient}$ radian is found, sample $n$ is deemed to be on an open edge that needs to be closed. Considering the blastholes are often sampled on a hexagonal lattice, we set $T_\text{orient}=\frac{2}{3}\pi$ and $K_\text{orient}=(4\times 2\pi/T_\text{orient})=12$. Some intermediate results are shown in Fig.~\ref{fig:cca-boundary-detection}.

One observation from Fig.~\ref{fig:cca-boundary-detection}(b) is that there is often a large gap between [what humans perceive as] adjacent samples along the [as yet undetermined] component boundaries. Indeed, the variable gap size renders useless boundary tracing algorithms like `marching squares' that rely on fixed separating distances. A more robust approach to boundary extraction is to formulate it as an energy functional minimization problem that provides a tradeoff between smoothness and fidelity. For this, projection of the boundary onto an image grid constitutes the first step.

\subsection{Edge map synthesis}\label{sect:edgemap-synthesis}
\begin{figure*}[!htb]
\centering
\includegraphics[width=156mm]{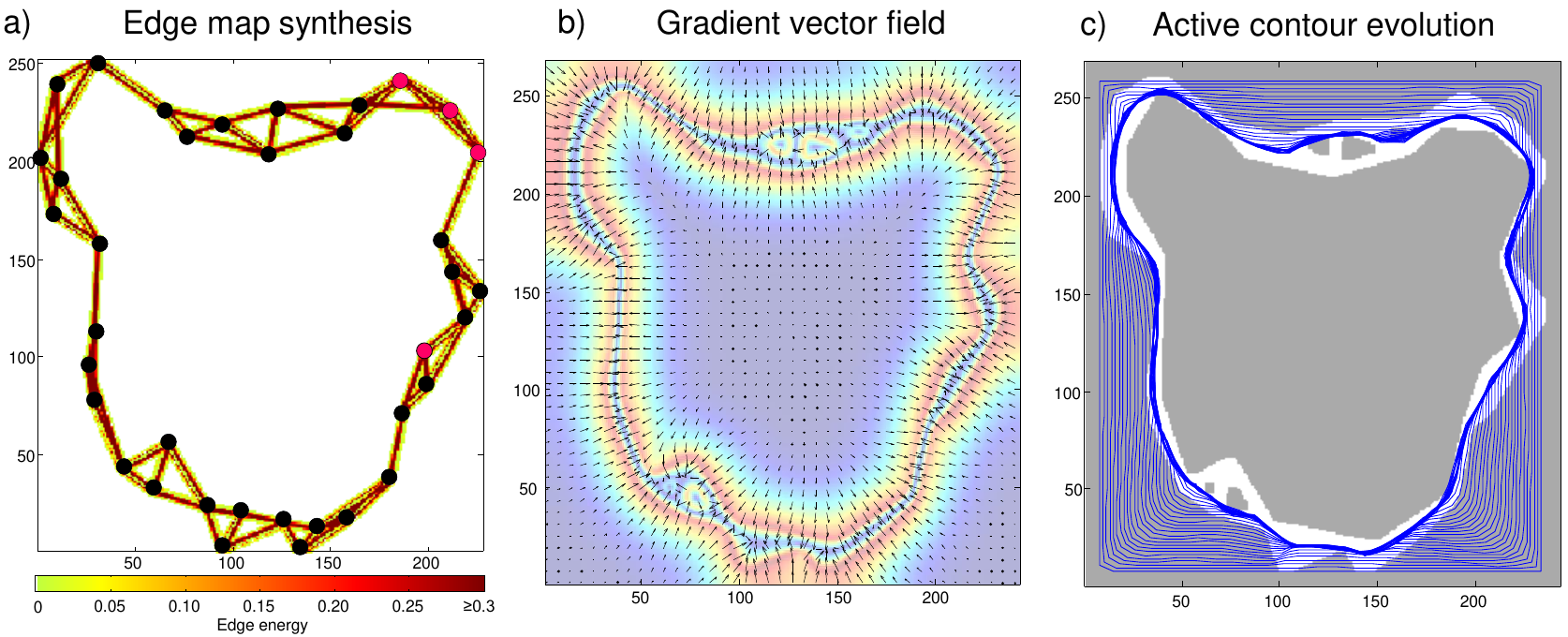}
\caption{(a) Synthesized edge map for a connected component after step 5 in Algorithm~2, boundary blastholes are overlaid. (b) Gradient vector field computed from the edge map. (c) Active contour evolves in the GVF, gravitating toward the region boundary at steady state. The default parameters are $K_\text{struct}=5$, $K_\text{joints}=4$, $n_\text{x}\times n_\text{y}\approx 240^2$, $p=8$.}
\label{fig:synthesis-gvf-ace}
\end{figure*}

The main objective is to bridge the gap between adjacent samples located on the boundary. Each sample connects with its $K$ nearest neighbors by forming an edge between them. This edge structure comprising of line segments is then densely sampled and transferred over to the image plane. Specifically, edge energy is accumulated by pixels within some margin of each line segment.\footnote{Morphological closing is subsequently applied to give the edges adequate thickness.} This procedure is further described in Algorithm 2.\footnote{Mapping real-world coordinates to the image domain (and vice-versa) involves scale changes. Uniform quantization is used implicitly throughout.} For illustration, a synthesized edge map is shown in Fig.~\ref{fig:synthesis-gvf-ace}(a). Although it contains some false edges, the active-contour segmentation approach (to be described next) can tolerate small imperfection.

\subsection{Active contour evolution in gradient vector field}\label{sect:bgm-active-contours-gvf}
The final goal is to obtain a contour (polygon of regularized boundary points) given the synthesized edge structure. This is achieved by performing region segmentation using Active Contour Evolution (ACE) in a Gradient Vector Field (GVF).  The GVF is obtained from the synthesized edge structure following the procedure given in Appendix~A. It may be visualized as a field of arrows (an external force) that pushes any given point in space toward the boundary (see Fig.~\ref{fig:synthesis-gvf-ace}(b)). Accordingly, an active contour which has been initialized as the component bounding box will evolve over time under the action of the GVF and be drawn inward until it converges at the region's boundary. This is illustrated in Fig.~\ref{fig:synthesis-gvf-ace}(c) and the short video (see multimedia content). The computational aspects of GVF active contour evolution are described in Appendix~A. This highlights the first connection  with computational physics.

\section{Spatial correspondence}\label{sect:bgm-spatial-correspondence}
\begin{figure}[!htb]
\centering
\includegraphics[width=85mm]{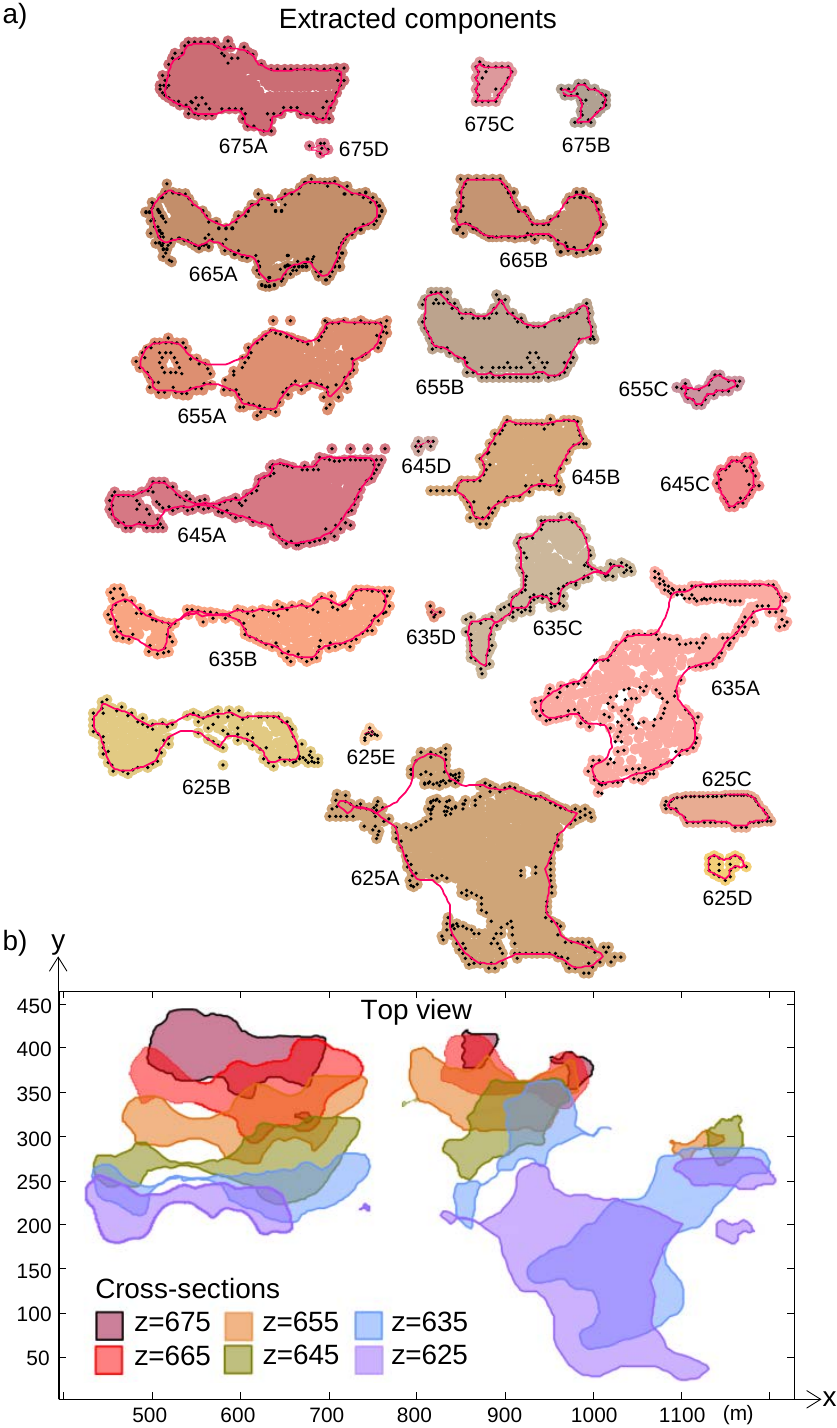}
\caption{Component boundaries extracted using active contours for six horizontal cross-sections (spaced 10m apart) in one geozone. Black dots denote blasthole samples located at\,/\,near geozone transitions. These serve as input to the next subsystem in the boundary geometry modeler called `spatial correspondence'. The goal is to associate regions and estimate component alignment between corresponding source and target contours subject to spatial constraints.}
\label{fig:extracted-components}
\end{figure}

The \textit{boundary extraction} subsystem produces a collection of contours. For instance, a set of real component boundaries extracted from multiple cross-sections can be seen in Fig.~\ref{fig:extracted-components}. When given two successive cross-sections, we will henceforth refer to the upper and lower cross-sections (likewise for the components contained within them) as `source' and 'target' respectively. This section presents general strategies for establishing relationships between source and target contours (viz., \textit{region association}) and estimating the optimal translation (performing \textit{component alignment}) with and without spatial constraints. The devised strategies take into account unique characteristics such as incompleteness, causality and resolution disparity in the data as foreshadowed in the introduction.

In this paper, spatial correspondence is treated as a two-stage process. The spatial mapping is the result of combining component translation and contour metamorphosis (local surface deformation). The present goal is to account for the rigid movement of associated components --- principally the translation observed in successive cross-sections. Subsequently, non-rigid movement will be considered in Section~\ref{sect:bgm-contour-metamorphosis}.

\subsection{Region association}\label{sect:bgm-region-association}
Given two cross-sections with $n_\text{s}$ source components and $n_\text{t}$ target components, the goal is compute the association matrix $A\in\mathbb{R}^{n_\text{s}\times n_\text{t}}$ which describes the relationships between $s$ and $t$. In the preliminary assessment, a `1' in $A(s,t)$ implies there is potential interaction between source $s$ and target region $t$.

\subsubsection{Formulation}
A Gaussian pdf $p_s(\mathbf{x}\!\mid\!\mathbf{d})\propto\exp\left(-a(\Vert\mathbf{x}-\mathbf{c}_s\Vert/b)^2\right)$ is used to model the likelihood of source-target association. In this setup, spatial proximity is measured by the difference of $\mathbf{x}$: an arbitrary location in the image and a fixed $\mathbf{c}_s$: the centroid position of source component $s$. Design choices revolve around the hyper-parameters $a$ and $b$ which must capture relevant aspects to produce sensible results. The right formulation very much depends on the problem, the sampling characteristics and reliability of the data.\footnote{Examples of probabilistic reasoning and state estimation techniques are plentiful in the literature. Those that incorporate geometry information for visual tracking \cite{koller-94,isard-98,rasmussen-01,tsechpenakis-04,williams-05,stenger-06,rathi-07} generally rely on having rich and accurate information delivered at a high frame rate. e.g., solving spatial correspondence problems using optical flow or persistent templates. In our application, shape invariance is not something that can be reliably exploited. Topological changes, under-sampling (large spacing between successive cross-sections) and sampling uncertainty (mixture of material and poor resolution) often contribute to sudden and unexpected changes. In geology and mining exploration, operational constraints also limit the amount of observations that can be gathered, this creates a situation where fast changing dynamics may surpass the learning rate for a complex model. Given the few samples available, we resort to using the most basic measures like the magnitude, but not the directionality, of motion.}

\textit{Apparent motion} and \textit{object area} represent two factors most relevant to our model. Apparent motion is inferred from the statistics $\mu_{\mathbf{d}_{\min}}$ and $\sigma_{\mathbf{d}_{\min}}$ which are computed from observed distances between the sources and targets, $d_{s,t}$ as follows.
\begin{align}
\mathbf{d}_{\min}&=\left[\min_t d_{s,t}\right]_s\in\mathbb{R}^{n_\text{s}\times 1},\\ 
\mu_{\mathbf{d}_{\min}}&=\mathbb{E}[\mathbf{d}_{\min}],\quad \sigma_{\mathbf{d}_{\min}}=\sqrt{\mathbb{E}[\mathbf{d}_{\min}^2]-\mu_{\mathbf{d}_{\min}}^2}
\end{align}
Variability in the observations is given by the standard error $s_{\mathbf{d}_{\min}}=\sigma_{\mathbf{d}_{\min}}/\sqrt{n_\text{s}}$. The association probability takes the form
\begin{align}
p_s(\mathbf{x}\!\mid\!\mathbf{d}_{\min})\propto \exp\left(-\dfrac{(\Vert\mathbf{x}-\mathbf{c}_s\Vert / \mu_{\mathbf{d}_{\min}})^2}{2\nu^2}\right),\\ \nu=\dfrac{4}{1+\exp(-s_{\mathbf{d}_{\min}}/s_0)}
\end{align}
The term $s_{\mathbf{d}_{\min}}/s_0$ relates to the uncertainty of the observations and is normalized by a reference value $s_0$. The `noise' parameter $\nu\in [2,4]$ regulates the shape of the Gaussian. When variability is low ($s_{\mathbf{d}_{\min}}/s_0\rightarrow 0$) the association function contracts and becomes more concentrated around $\mathbf{c}_s$. Conversely, it dilates when there is a lack of consensus on $\mathbb{E}[\mathbf{d}_{\min}]$; this allows more distant targets to become potentially associated with the source.

\subsubsection{Scale effects}
Since the $\mathbf{d}_{\min}$ observations are noisy, component size should be compared with the apparent translation $\mu_{\mathbf{d}_{\min}}$. When $\mu_{\mathbf{d}_{\min}}$ exceeds the object span $\varphi_s$, the policy $\mu_{\mathbf{d}_{\min}}\leftarrow \min\{\,\varphi_s,\mu_{\mathbf{d}_{\min}}\}$ should be adopted. This limits the scope of association for smaller components by shrinking $p_s(\mathbf{x}\!\mid\!\mathbf{d}_{\min})$.\footnote{The object span is approximated from the area as $\varphi_s\approx \sqrt{a_s/\pi}$.} To prevent $p_s(\mathbf{x}\!\mid\!\mathbf{d}_{\min})$ from becoming too \textit{narrow} --- which would eliminate any source-target association potential when taken to the extreme --- it would be prudent to also place a lower limit $(\mu_\text{lower})$ on $\mu_{\mathbf{d}_{\min}}$.

The final form of the association likelihood is given by
\begin{align}
p_s(\mathbf{x}\!\mid\!\mathbf{d}_{\min})&\propto\exp\left(-\dfrac{(\Vert\mathbf{x}-\mathbf{c}_s\Vert/\lambda_s)^2}{2\nu^2}\right),\ \mathbf{x},\mathbf{c}_s\in\mathbb{R}^2\label{eq:assoc-probability}\\
\lambda_s&=\max\{\,\min\{\,\mu_{\mathbf{d}_{\min}},\varphi_s\},\,\mu_\text{lower}\}\label{eq:assoc-prob-lambda}\\
\nu&=\dfrac{4}{1+\exp(-s_{\mathbf{d}_{\min}}/\mu_\text{lower})}\label{eq:assoc-prob-nu}
\end{align}
where $\mu_\text{lower}$ is related to the anticipated lateral movement. It basically dictates the minimum radius of association.

\subsubsection{Decision function}
The decision of whether to associate source $s$ with target $t$ is governed by a discriminant function
\begin{align}
A(s,t)&=\begin{cases}\,1 & \text{if percentile}\!\left(\{p_s(\mathbf{x}\!\mid\!\mathbf{d}_{\min})\!\mid\! \mathbf{x}\!\in\!\mathcal{R}_t\},K\right)\ge 0.5\\
\,0 & \text{otherwise}
\end{cases}\label{eq:discriminant-percentile}\\
K&=50\times\left(1+\left(1-\min\left\{\frac{a_s}{a_t},1\right\}\right)\right)\in[50,100] \label{eq:area-ratio-requirement}
\end{align}
\begin{itemize}
\item The set notation $\{p_s(\mathbf{x}\!\mid\!\mathbf{d}_{\min})\mid \mathbf{x}\in\mathcal{R}_t\}$ in (\ref{eq:discriminant-percentile}) refers only to points inside the support interval of target component $t$.
\item The percentile $K$ in (\ref{eq:area-ratio-requirement}) depends on the source-to-target area ratio ($a_s/a_t$). The rationale is as follows:
  \begin{itemize}
  \item If $a_s < a_t$, the source can (at best) fit itself wholly within the target with room to spare. Thus, a fair value for assessing spatial proximity (the association potential) is the median likelihood, offset by the fraction of $a_t$ which $s$ cannot overlap.
  \item In the limit as $(a_s/a_t)\rightarrow 0$, $K$ seeks out the maximum likelihood when $s$ is treated as a point object.
  \end{itemize}
\end{itemize}

\subsection{Component alignment strategies}\label{sect:bgm-component-alignment-strategies}
\begin{figure}[!htb]
\centering
\includegraphics[width=84mm]{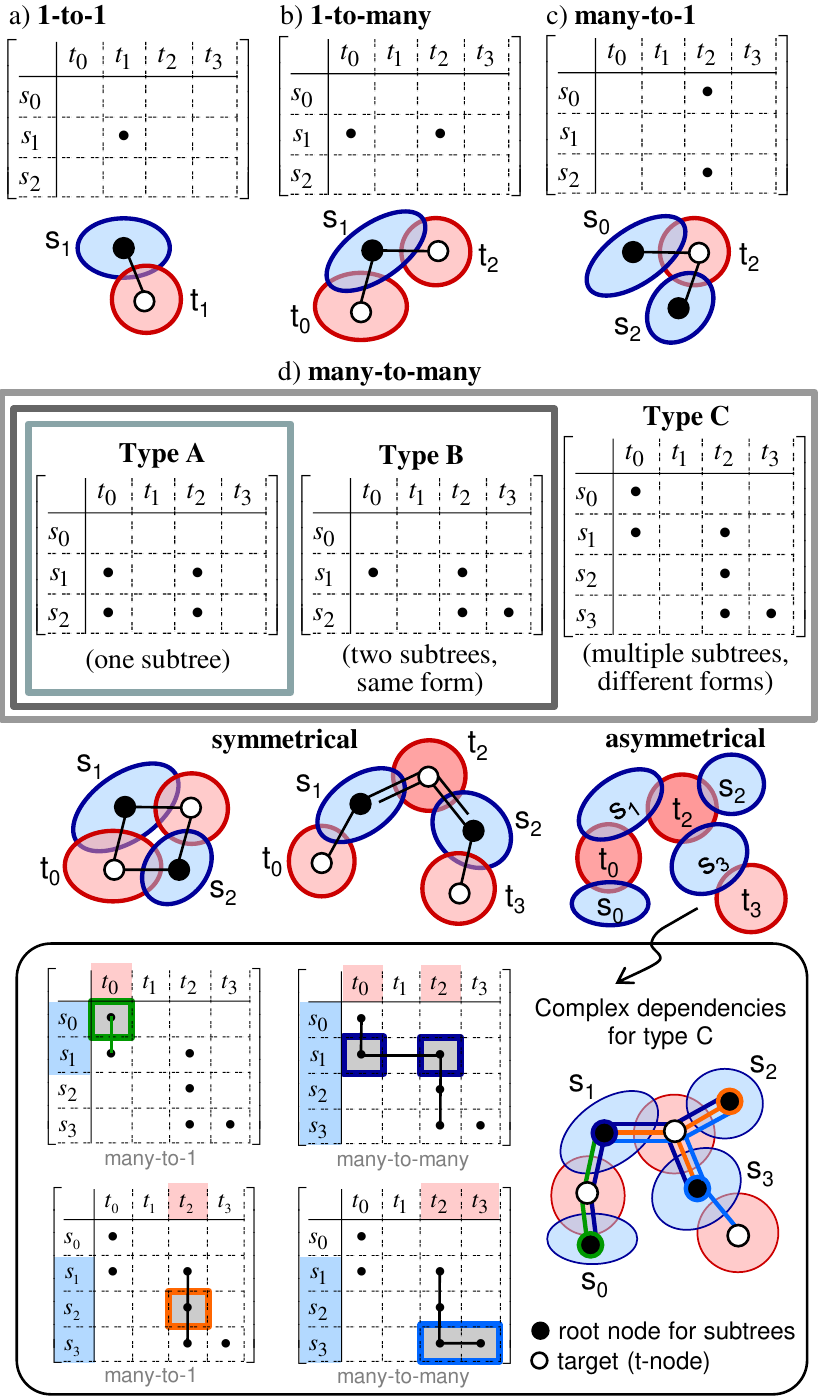}
\caption{Source--target component relationships (viz., pattern of potential overlap) are represented by intersection graphs given the association matrix. Source and target components are denoted by black and white nodes respectively. Different region association scenarios such as multiple-source single-target are depicted in (a)--(d). For complex scenarios (see inset), suitable component alignment strategies will be applied to each subtree anchored at an s-node which includes all connected t-nodes and their children s-nodes.}
\label{fig:graphical-decomposition}
\end{figure}

Having obtained the association matrix $A(s,t)$, source--target relations can be represented by a graph. This gives rise to several possibilities: a 1-to-1 mapping, many-to-one, one-to-many, and many-to-many correspondence between the source and target components. These scenarios are shown in Fig.~\ref{fig:graphical-decomposition}. For complex scenarios (e.g., multiple sources associated with multiple targets), the intersection graph is decomposed into subtrees (see Fig.~\ref{fig:graphical-decomposition} inset). This section is concerned with devising suitable strategies for each scenario, i.e., finding source translation estimates which maximize target alignment or component coverage. Graphically, each scenario is described by a subtree rooted at an s-node and its expansion includes all connected t-nodes and their children (any s-nodes connected to the t-nodes).

\subsubsection{Displacement estimation: cross-correlation via FFT}\label{sect:bgm-displacement-estimation-fft}
A 1-to-1 mapping, or single-source single-target alignment, is the simplest situation encountered in region association. Conceptually, the best alignment is achieved by template matching; this is otherwise known as 2D cross-correlation. Using vectorized notation, e.g., $\mathbf{i}=(i_0,i_1)$, the problem may be posed as
\begin{align}
\text{optimal translation } = \argmax_{\mathbf{m}}r_{VU}[\mathbf{m}]\label{eq:optimal-translation}
\end{align}
where the matched filter response with lag $\mathbf{m}$ is given by
\begin{align}
r_{VU}[\mathbf{m}]\!=\!\sum_\mathbf{i} v[\mathbf{i}]\,u[\mathbf{i\!+\!m}]\!=\!\sum_{i_0,\,i_1}\!v[i_0,i_1]\,u[i_0\!+\!m_0,i_1\!+\!m_1]\label{eq:rxy-definition}
\end{align}

These operations can be implemented efficiently as
\begin{align}
r_{VU}[\mathbf{m}]= \mathcal{F}^{-1}(V[\mathbf{k}]\ U^{*}[\mathbf{k}])\label{eq:peak-via-fft}
\end{align}
using the Fast Fourier Transform (separable FFT) \cite{gonzalez-woods-02}. In this equation,\footnote{Caveats: Certain technical conditions need to be satisfied for this to work. Stated simply, the inputs need to have first quadrant spatial support and be sufficiently padded around the margins to ensure no shifted component ever steps outside the image border. Detailed guidance is given in \cite{gonzalez-woods-02} \S4.6.4.} $r_{VU}$ represents the 2D cross-correlation between $u$ and $v$, where $u\equiv\mathcal{U}$ is the source mask (a binary indicator function for the region occupied by the source component in the image plane) and $v\equiv\mathcal{V}$ is a signed distance function \cite{felzenszwalb-12} (SDF) for the target component. The SDF (computed in Algorithm~3) gives a 2D map of the signed distance of any location from the contour boundary (zero-interface) and adheres to the convention where pixels inside and outside the target region have positive and negative values, respectively.

On the RHS, $\mathcal{F}^{-1}$ represents the inverse FFT. $U^{*}[\mathbf{k}]$ and $V[\mathbf{k}]$ denote the complex conjugate of the FFT coefficients for $u$ and FFT coefficients for $v$. Finding the optimal shift (displacement $\mathbf{m}$) is equivalent to locating the peak in $r_{VU}$. This lays the basic foundation for component alignment (see Algorithm~4).

\subsubsection{Multiple-source single-target scenario}\label{sect:bgm-many-to-1-scenario}
\begin{figure}[!htb]
\centering
\includegraphics[width=165mm]{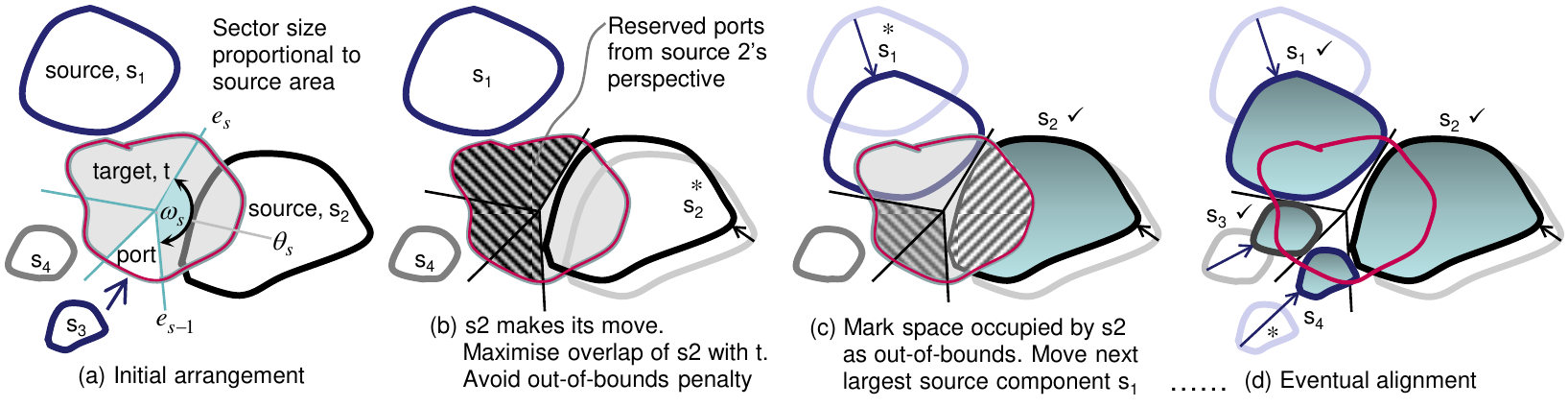}
\caption{Port allocation concept for multiple-source single-target scenario}
\label{fig:port-allocation-for-many-to-one}
\end{figure}

The proposed solution (\ref{eq:optimal-translation}) requires certain modifications in a multiple-source single-target (many-to-one) scenario. Consider the situation depicted in Fig.~\ref{fig:port-allocation-for-many-to-one}(a). As it stands, if source $s_2$ acts selfishly and tries to maximize its correlation with the target with no regard for other source components, $s_1$, $s_3$ and $s_4$ will be locked out of the region merging arrangement. To prevent this, the sources will participate in target sharing. This is posed as a \textit{resource allocation} problem where the objective is to divide the target into sectors (or ports of varying sizes) and assign the ports in such a way as to minimize their angular difference with the affiliated source component (see Fig.~\ref{fig:port-allocation-for-many-to-one}(a)). The solution is presented in Algorithms 5--6. The port boundaries and target sector-to-source mapping so obtained allow very specific spatial constraints to be imposed. For instance, in Fig.~\ref{fig:port-allocation-for-many-to-one}(b), by imposing a penalty on target sectors reserved for other source components (see striped region), $s_2$ will refrain from stepping into other's territory whilst maximizing its overlap with the target. This behavior can be realized by setting $v[\mathbf{i}]$ (initially, the target signed distance function) to large negative values at locations $\mathbf{i}\equiv(i_0,i_1)$ which are marked out-of-bounds.

\subsubsection{Single-source multiple-target scenario}\label{sect:bgm-1-to-many-scenario}
\begin{figure}[!htb]
\centering
\includegraphics[width=165mm]{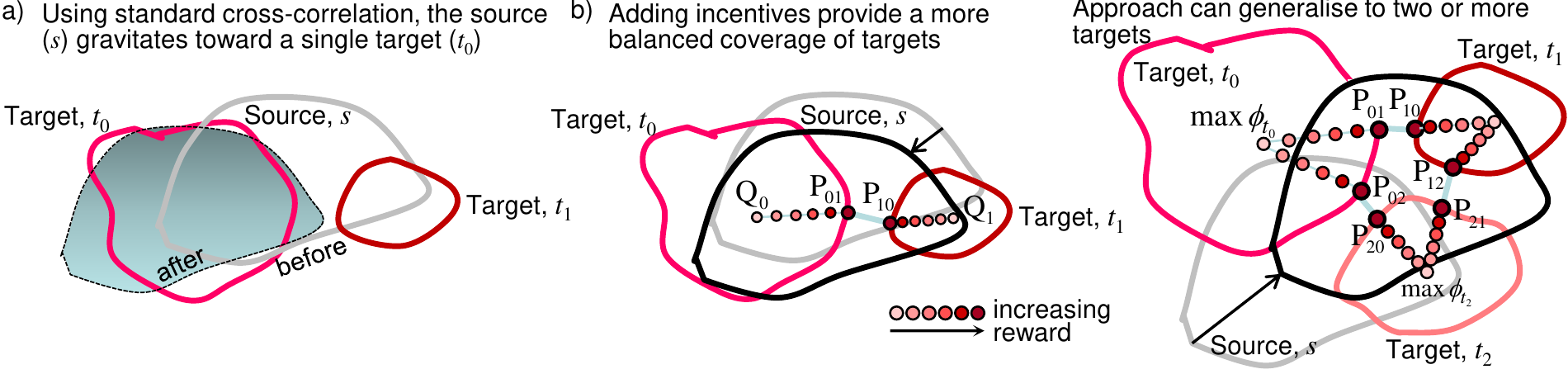}
\caption{Target coverage in single-source multiple-target scenario}
\label{fig:incentives-for-one-to-many}
\end{figure}

An issue with the current approach is that it is somewhat biased toward single-target alignment. There is no \textit{incentive} for a source component to overlap with multiple targets since stepping outside the boundary of a target incurs a negative penalty in (\ref{eq:rxy-definition}). This usually deters any `cross-over' unless the source component is large enough to encompass both targets and the targets are reasonably close together that reward outweighs penalty. As motivation, Fig.~\ref{fig:incentives-for-one-to-many}(b) illustrates a situation where the source is encouraged to straddle two targets; this in turn maximizes component coverage and offers a more plausible explanation in a region splitting scenario. Within the current framework, this can be achieved by placing rewards at strategic locations in the value function $v$ to promote cross-over. This is further described in Algorithm 7.

\subsubsection{Multiple-source multiple-target scenario}\label{sect:bgm-many-to-many-scenario}
\begin{figure}[!thb]
\centering
\includegraphics[width=85mm]{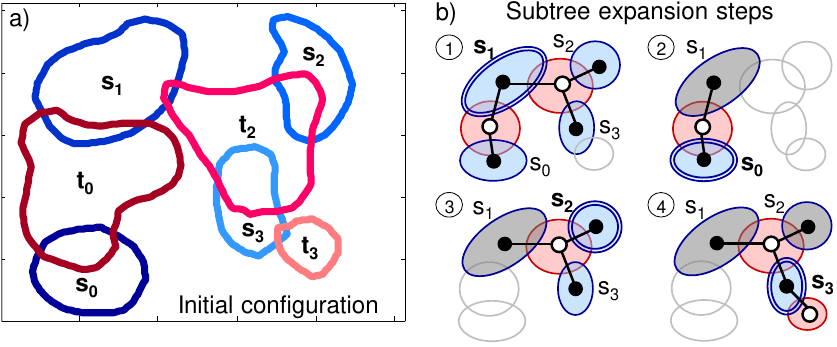}
\caption{Spatial correspondence in multiple-source multi-target scenario. Refer to the \href{https://ieeexplore.ieee.org/ielx7/6287639/8600701/8891690/supplementalmaterial.pdf?tp=&arnumber=8891690}{Supplementary Material} for a worked example with details of the expansion steps.}
\label{fig:integrated-steps-many-to-many-preview}
\end{figure}

The \textit{cooperative} and \textit{reward} strategies described in \S\ref{sect:bgm-many-to-1-scenario}--\ref{sect:bgm-1-to-many-scenario} provide a unified way for dealing with complex relationships under the displacement estimation framework of \S\ref{sect:bgm-displacement-estimation-fft}. As an illustration, these techniques are applied to a multiple-source multiple-target scenario in 
Fig.~\ref{fig:integrated-steps-many-to-many-preview}. The initial position of the source and target components are shown in Fig.~\ref{fig:integrated-steps-many-to-many-preview}(a). This has the same graphical structure as the type C configuration in Fig.~\ref{fig:graphical-decomposition}. Hence, it can be decomposed into 4 subtrees or expansion steps as shown in Fig.~\ref{fig:integrated-steps-many-to-many-preview}(b).

A compact way of writing these steps is {\small\circled{1}}: $\mathbf{s_1}\!\rightarrow\! \{t_0\!\leftarrow\!s_0, t_2\!\leftarrow\!(s_2,s_3)\}$,\ {\small\circled{2}}: $\mathbf{s_0}\!\rightarrow\!t_0\!\leftarrow\!s_1$, {\small\circled{3}}: $\mathbf{s_2}\!\rightarrow\!t_2\!\leftarrow\!(s_1,s_3)$, {\small\circled{4}}: $\mathbf{s_3}\!\rightarrow\!\{t_2\!\leftarrow\!(s_1,s_2),t_3\}$ where the source component under consideration (to be moved) is typed in bold.

Within a subtree, each link describes a tentative association between an s-node and a t-node in Fig.~\ref{fig:integrated-steps-many-to-many-preview}(b). This simply means there is a potential for the source to intersect with the target. The objective is to eliminate edges where a connection does not exist --- perhaps due to obstacles (the presence of other source components) which have prevented an overlap between $s$ and $t$ --- and find the translation for source components that do overlap with targets (i.e., maximize correlation subject to spatial constraints).

For brevity, the subtree expansion steps are omitted here. Interested readers may refer to the worked example given in the \href{https://ieeexplore.ieee.org/ielx7/6287639/8600701/8891690/supplementalmaterial.pdf?tp=&arnumber=8891690}{Supplementary Material} for a full description of the expansion steps. Through the lens of spatial reasoning (Algorithm~3--9 and Figure~\ref{fig:synthesis-gvf-ace}--\ref{fig:integrated-steps-many-to-many-preview}), Section~\ref{sect:bgm-spatial-correspondence} has outlined the main connections with signal processing, computer vision, resource allocation and graphical decomposition in this work.

In summary, the \textit{spatial correspondence} subsystem produces an association matrix $A$ and translation estimates $\mathcal{T}$ which describe the relationship between source and target components in two successive cross-sections. Specifically, the source contours $\{C_s\}$ associated with a given target $C_t$ define an instance for the next subsystem, \textit{contour metamorphosis}, to work on.

\section{Contour metamorphosis}\label{sect:bgm-contour-metamorphosis}

\begin{figure*}[htb]
\centering
\includegraphics[width=165mm]{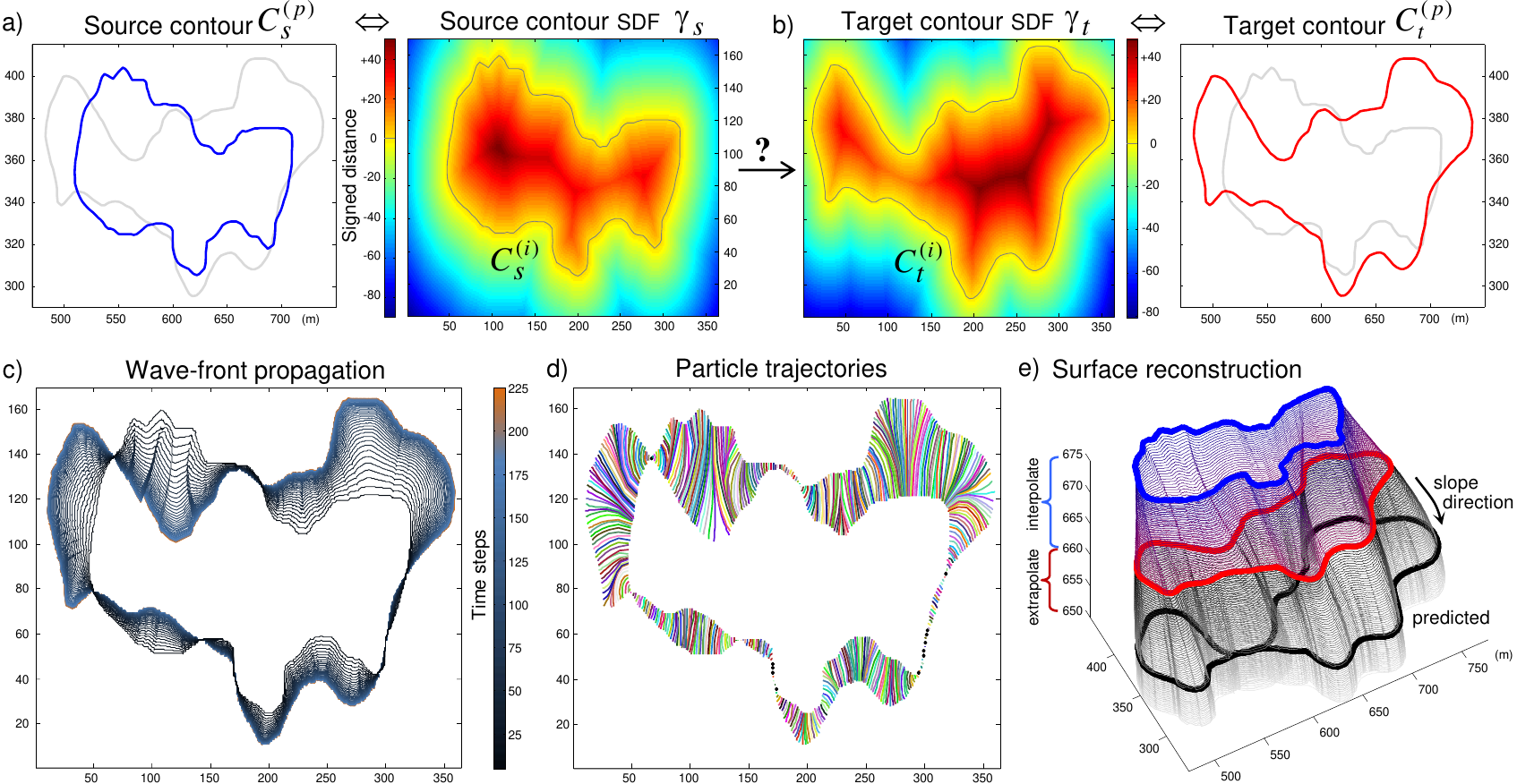}
\caption{Overview of contour metamorphosis. Basic objective is to model how the source region morphs into the target region using differential geometry, viz., explain how (a) is transformed into (b) using the narrative of (c) and (d) to ultimately produce a surface model similar to (e). Notation-wise, superscripts $(p)$ and $(i)$ denote the physical domain and image processing (modeling) domain respectively. Physical coordinates, e.g. $C^{(p)}_s=(\mathbf{x}^{(p)}_s,\mathbf{y}^{(p)}_s)$, are measured true-to-scale and expressed in UTM coordinates in [m]. (a) A source contour (blue line: $C_s^{(p)}$) is aligned with the target contour (light gray) in a common frame, it has an equivalent level-set representation $\gamma_s$ where contour boundary is embedded as the zero-interface. (b) Associated target contour (red line: $C_t^{(p)}$) and its level-set (signed distance) function $\gamma_t$. The arrow annotated with ``?'' is elaborated in (c) which explains the evolution process as a propagating wavefront, viz., the movement of the zero-interface as (a) morphs into (b). In terms of tracking, (d) provides an alternative description using particle trajectories which emphasizes on directionality.}
\label{fig:contour-metamorphosis-overview}
\end{figure*}

The objective is to capture local spatial variation using differential geometry, to explain how source regions evolve into target regions using PDE. This problem may be visualized as a transformation from (a) to (b) in Fig.~\ref{fig:contour-metamorphosis-overview}.

Given displacement estimates $\{\mathbf{m}_s=(m^{(x)}_s,m^{(y)}_s)\}_s$ and translation operator $\mathcal{T}(\mathbf{x},\mathbf{y})=(\mathbf{x}+m^{(x)}_s,\mathbf{y}+m^{(y)}_s)$, the starting point is a principal target contour $C_t\equiv(\mathbf{x}_t,\mathbf{y}_t)$, its associated (shift-compensated) source contours $\{C'_s\!\equiv\!\mathcal{T}(\mathbf{x}_s,\mathbf{y}_s)\}$ and any affiliated target contours $\{C_{t'}\}$ connected to the source nodes. Fig.~\ref{fig:contour-metamorphosis-overview}(a) shows a simple case where there is only one source and one target contour involved. It is this residual difference, the non-rigid changes in shape in the aligned common frame, that is being modeled.

In the PDE model, contours are represented by level-sets $\phi_t$ which embed region boundaries as the zero-interface, viz., $\Omega_t=\{\mathbf{n}\in\mathbb{R}^2\mid|\phi_t[\mathbf{n}]|\le 0.5\}$. Initially, $\phi_{t=0}$ is set to $\gamma_s$, the signed distance function of $C_s$. The basic objective is to model the evolutionary process from $\phi_{t=0}\!=\!\gamma_s$ to $\phi_{t=\infty}\!=\!\gamma_t$. As the level-set $\phi(t)$ evolves\footnote{A level-set may be defined as a collection of iso-contours or a family of embedded point sets each represented by $\mathcal{S}_k=\{\,\mathbf{n}\mid\phi(\mathbf{n})\ge k\,\}$.}, the region boundary morphs from (a) to (b). The morphing process is described in Appendix~B wherein (20) represents the key equation for level-set update. Visually, this computation describes a \textit{propagating wavefront} as depicted in Fig.~\ref{fig:contour-metamorphosis-overview}(c).

\subsection{Iso-contours and wave speed adjustment}
It is important to note that these level-set zero interfaces $\Omega_t$ are functions of time. Ultimately, we need to obtain iso-contours at different elevations $z$. However, no single $\Omega_t$ (other than $\Omega_0$) currently corresponds to a fixed elevation.

As a corollary, not all points on the advancing wavefront will reach the target contour at the same time since the wave propagates at different speeds depending on location.\footnote{The propagation speed depends on the geometric distance between the interface and target contour in accordance with (19).} Therefore, \textit{particle trajectories} (see Fig.~\ref{fig:contour-metamorphosis-overview}(d)) are used to track the direction in which the zero-interface moves to facilitate slope estimation at a specific height. The relevant procedure called ``particle advection'' is described in Appendix~B wherein (26) represents the key equation for particle update.

The particle speed (spacing) along each trajectory will eventually be adjusted to ensure each advances at a uniform rate and reaches the target boundary at the same time. This allows the zero interfaces to be converted into iso-contours of constant elevation. These ideas capture the essence of contour metamorphosis; Fig.~\ref{fig:contour-metamorphosis-overview}(e) also serves as a prelude of the end goal.

\subsection{Challenges}\label{sect:tracking-challenges}

\begin{figure*}[ht]
\centering
\includegraphics[width=165mm]{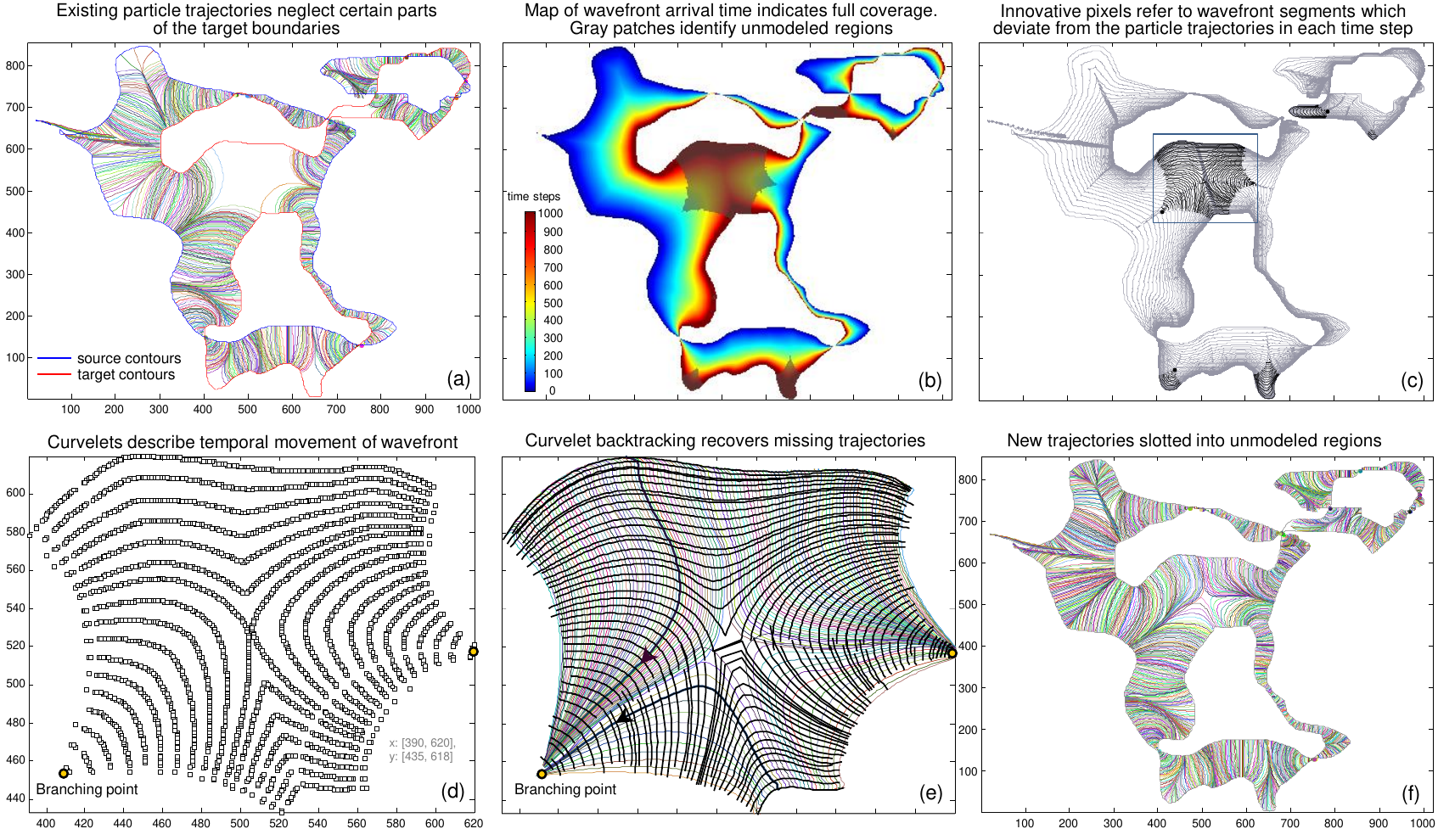}
\caption{Illustration of branching point phenomenon. (a) Particle trajectories (obtained using Appendix~B) reveal target segments with zero particle coverage (see empty red lines). (b) Time of arrival map indicates the target segments concerned were clearly reached by the wavefront, gray area indicates region with no trajectory coverage. (c) Innovative areas where the wavefront diverges from the evolving particle trajectories are identified by black lines in each time step. (d) Magnified view of the window in (c) demonstrating the concept of curvelets, a collection of pixels (only a subset is shown), which describe the temporal movement of the wavefront, specifically, the portions which the particles have lost track of. These emanate from a single branching point. (e) Curvelet back-tracking between successive time steps as a means of re-establishing the missing trajectories. Final result: (f) Merging the recovered curvelet trajectories with existing particle trajectories.}
\label{fig:branching-problem-overview}
\end{figure*}

In principle, Appendix~B provides all the necessary tools for capturing information on local spatial deformation. In practice, when the shape of the source and target regions are dissimilar, the scheme may fail. Specifically, particle trajectories may fail to track the wavefront in areas where complex topological changes occur especially when they coincide with large differences in local curvature. This can be seen in Fig.~\ref{fig:branching-problem-overview}(a) where some target boundary segments have zero particle coverage. Fig.~\ref{fig:branching-problem-overview}(b) highlights regions which end up being unmodeled despite being reached by the wavefront. This usually occurs due to undersampling when the cross-section spacing is large. The goal in this section is to determine its root cause and contribute a solution to this problem.

In each time step, wavefront segments which the particle trajectories had failed to track are archived. This produces a collection of time-stamped pixels (called \textbf{curvelets}) which identify areas of innovation, viz., locations where the wavefront and head of the particle trajectories have diverged. The window of interest in Fig.~\ref{fig:branching-problem-overview}(c) is magnified in (d) to reveal its structure. It can be seen the waves emanate from two separate sources and merge in the middle, this describes a region splitting situation which can be reasonably expected based on the topology shown in (a). For other unmodeled regions shown in (c), the situation is less complex and the behavior can be attributed to significant differences in local curvature. The common theme is that the missing particle trajectories appear to emerge from a single \textbf{branching point}.

The color strands in Fig.~\ref{fig:branching-problem-overview}(e) show the trajectory flow for new particles sitting on the curvelets. The arrows indicate the direction the \textbf{backtracking} algorithm proceeds in. The aim is to establish pathways from the target back toward the branching point. In Fig.~\ref{fig:branching-problem-overview}(f), new pathways deduced from the curvelets are slotted into the unmodeled regions. Evidently, this recovers directional information between the source and target contours in areas where information had previously been lost.

\subsection{Curvelet backtracking}\label{sect:curvelet-backtracking}
\begin{figure}[ht]
\centering
\includegraphics[width=165mm,trim={0mm 0 0 0},clip]{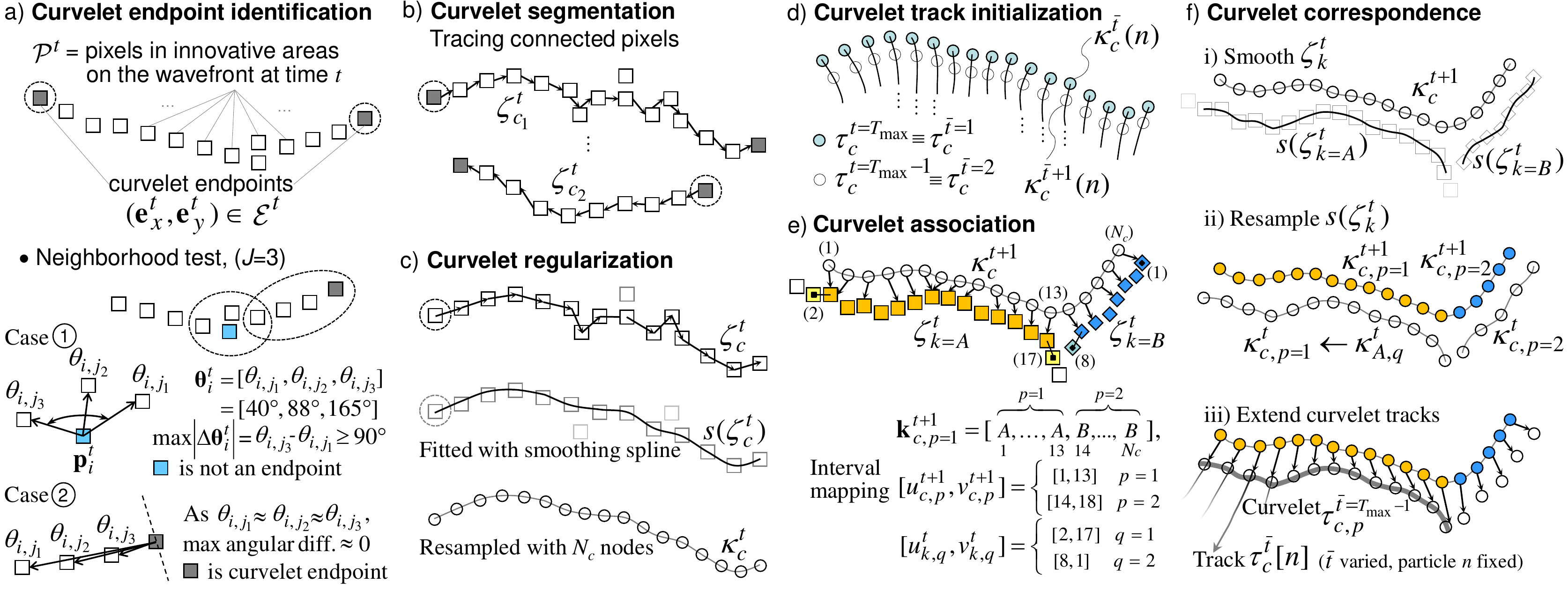}
\caption{Curvelet backtracking concepts (cf. Algorithm\,10)}
\label{fig:curvelets-illustrations}
\end{figure}

The backtracking algorithm is formally described in Algorithm~10. The starting point is a collection of time-stamped pixels, these represent wavefront segments the particle trajectories had failed to track using the particle update equation (26) during contour metamorphosis. Backtracking comprises six primitive steps; these are described below and illustrated in Fig.~\ref{fig:curvelets-illustrations}.
\begin{enumerate}[label=(\alph*)]
\item \textbf{Curvelet endpoint identification}\newline
Locate the start and endpoints of prospective curvelets within $\mathcal{P}^t$ --- the set of unordered pixels on the wavefront which have drifted away from the particle trajectories at time $t$.
\item \textbf{Curvelet segmentation (clustering and ordering)}\newline
Extract individual curvelets $\zeta^t_c$ by tracing the connecting pixels between relevant endpoints.
\item \textbf{Curvelet regularization}\newline
Fit pixels with a smoothing spline to remove jitter. Resample segment as $N_c$ point regularized curvelet $\kappa^t_c$.
\item \textbf{Curvelet track initialization\,/\,extension}\newline
The displacement of curvelet points between successive time steps $[\tau^{t}_c(n),\tau^{t-1}_c(n),\ldots]$ describe possible pathways for the missing trajectories in unmodeled regions.
\item \textbf{Curvelet association}\newline
Given regularized points $\kappa^{t+1}_c(m)$ for curvelet $c$ at time $t+1$, find a matching curvelet $k$ at time $t$.\footnote{Nearest neighbor search is employed to find a surjective mapping from $t$ to $t+1$. Each point $\kappa^{t+1}_c(m)$ has at least one match at $t$, but not every pixel on $\zeta^{t}_k$ is matched with a point at $t+1$. For this reason, as illustrated in  Fig.~\ref{fig:curvelets-illustrations}(e), this is used only for curvelet association and not for point-wise correspondence. This is because inconsistent (jittery) movement has a detrimental effect on trajectory estimation. To eliminate the cumulative effects of drifting which would otherwise skew the tracks over time, the associated point set is constrained to lie on a regularized curvelet as shown in Fig.~\ref{fig:curvelets-illustrations}(f).} A case of interest is depicted in Fig.~\ref{fig:curvelets-illustrations}(e) where the segmented pixels $\zeta^t_k(n)$ from two curvelets $k=A,B$ are merged from $t$ to $t+1$. This is handled by considering interval mappings $\mathbf{k}^{t+1}_{c,p}$ for multiple curvelet parts ($p=1,2$).
\item \textbf{Curvelet correspondence}\newline
Once matching pixels $\zeta^{t}_k$ are found, the curvelet at time $t$ is smoothed and resampled evenly with $N_c$ points. This curvelet $\kappa^{t}_k$ is then collectively mapped onto $\kappa^{t+1}_c$, the current curvelet.\footnote{This provides an interesting contrast to the wavefront tracking approach proposed by Tomek et al.\,\cite{tomek-16} which represents two sets of time-stamped pixels with a complete oriented bipartite graph, poses the task of finding the best tracking arrows as a network flow problem, formulates and solves it as an integer program using branch and bound. Similarly, Singh et al.\, treated this as an optimization problem and employed dynamic programming in \cite{singh-97}.} For a given point $n$, the sequence [$\kappa^{t+1}_c(n),\kappa^{t}_c(n),\kappa^{t-1}_c(n),\ldots$] subsequently defines a track which traverses the unmodeled region.
\end{enumerate}

\begin{figure}[ht]
\centering
\includegraphics[width=89mm,trim={1mm 0 0 0},clip]{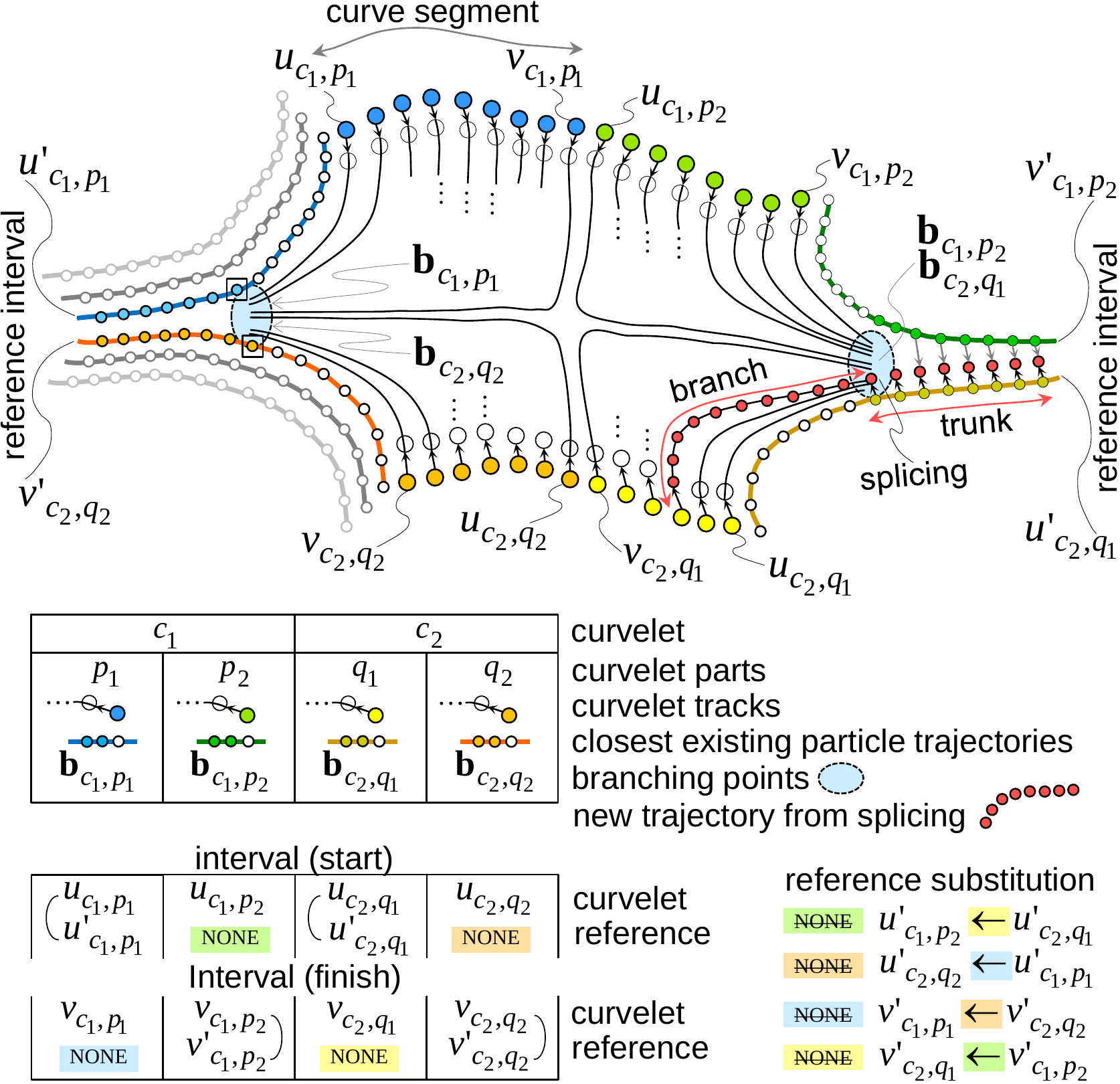}
\caption{Curvelet branching considerations (cf. Algorithm\,11--12)}
\label{fig:curvelets-branching-analysis}
\end{figure}

These steps capture the essence of the backtracking algorithm which produces particle tracks, see color strands in Fig.~\ref{fig:branching-problem-overview}(e) and (f), for the unmodeled regions. The remaining tasks involve branching point identification which enables existing particle trajectories [the trunk portion extending from the source contour to the branching point] to be fused with the newly discovered tracks [from the branching point to the target contour] to complete the missing pieces. The details are described in Algorithms 11 and 12 and the notations are clarified in Fig.~\ref{fig:curvelets-branching-analysis}.

\section{Results}\label{sect:results}
The proposed boundary extraction technique was applied to blasthole patterns from a Pilbara iron ore mine in Western Australia. Fig.~\ref{fig:triangulated-surfaces}(a) illustrates one such pattern for a mineralized geozone before any modeling was done.

\subsection{Visualization}\label{sect:visual}
\begin{figure*}[ht]
\centering
\includegraphics[width=165mm,trim={0 1mm 0 0},clip]{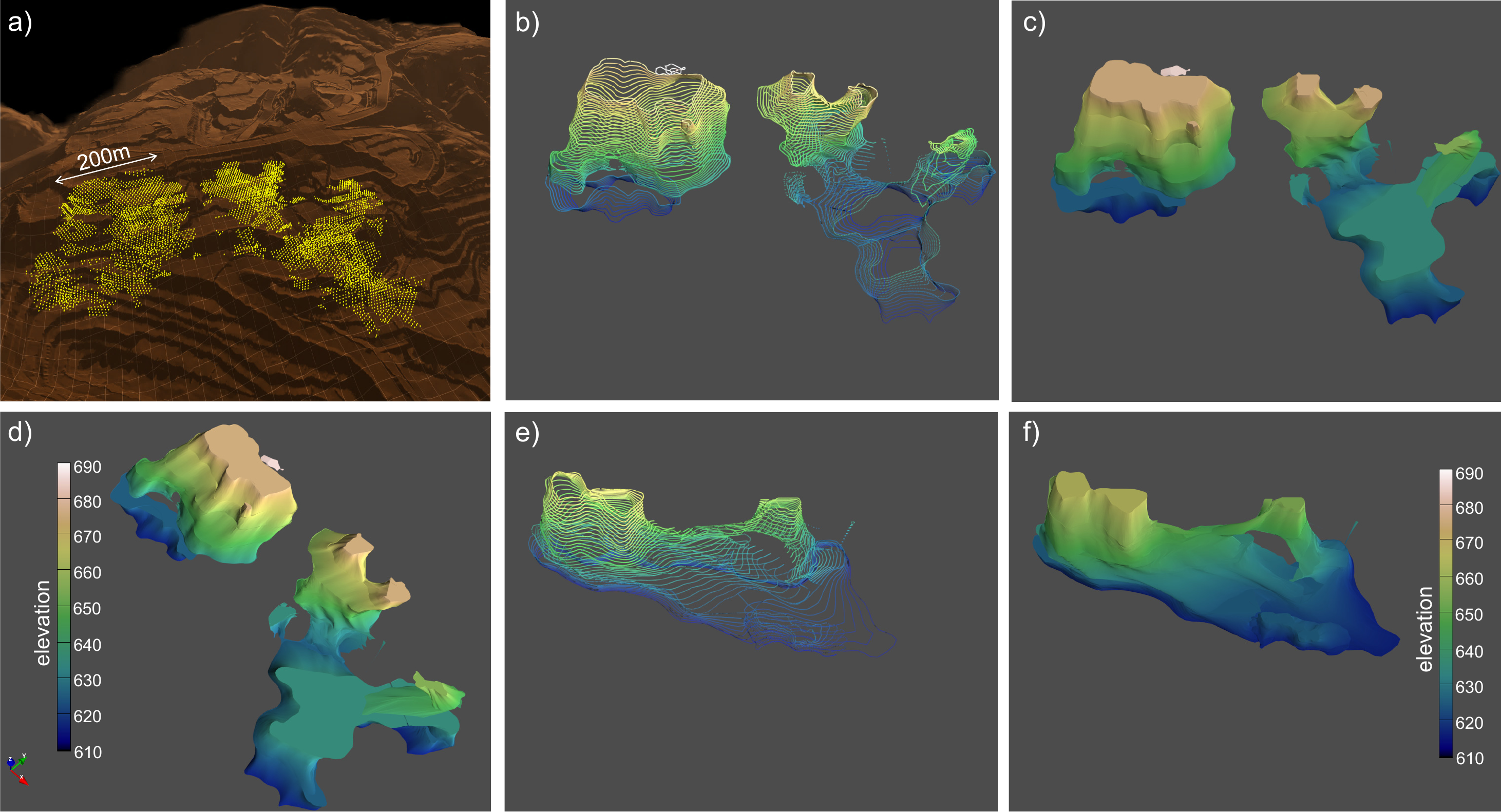}
\caption{(a) Blasthole pattern superimposed on the terrain. Extracted contours and triangulated surfaces for two geozones: $g_1$ in (b)-(c) and $g_2$ in (e)-(f). The surface appearing in (c) is shown from a different vantage point in (d).}
\label{fig:triangulated-surfaces}
\end{figure*}

\begin{figure}[ht]
\centering
\includegraphics[width=150mm]{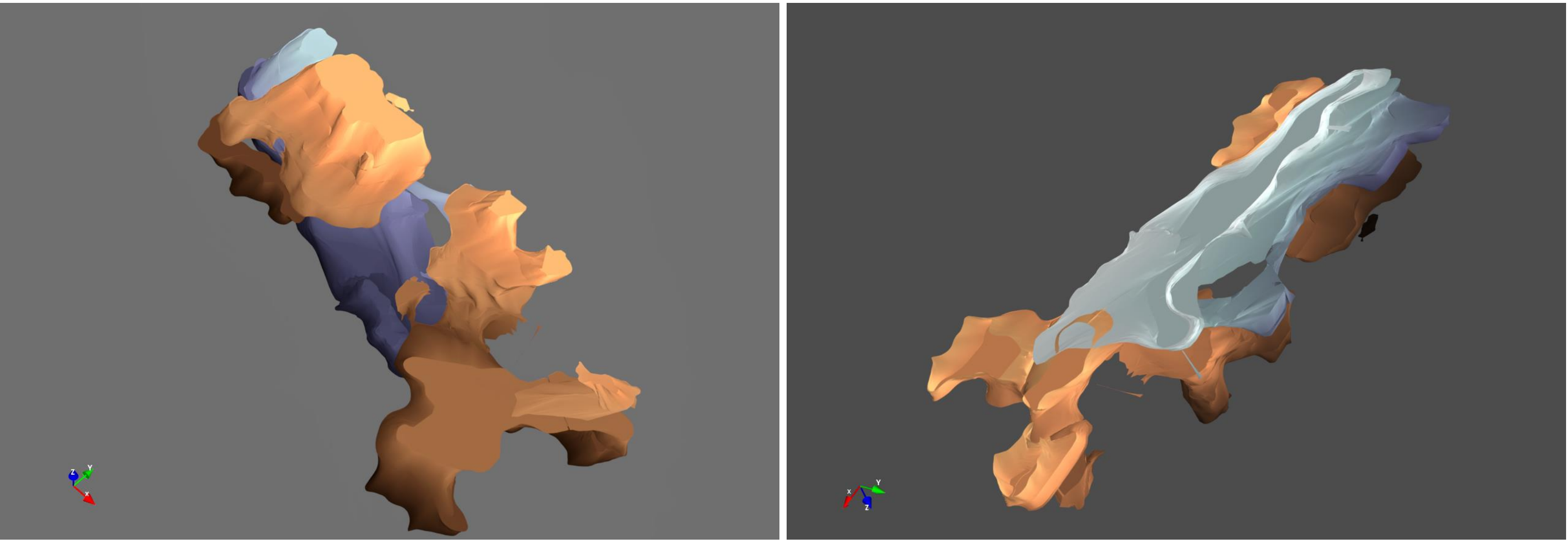}
\caption{Extracted surfaces for geozone $g_1$ (orange) and $g_2$ (blue) conform to the local geological structure where one is stacked on top of the other. Two vantage points from above and below in the top and bottom panels.}
\label{fig:bedded-surfaces}
\end{figure}

Spatial correspondence and metamorphosis were applied to the segmented regions; culminating in the contours seen in Fig.~\ref{fig:triangulated-surfaces}\,(b and e). These contours are triangulated to produce the surfaces shown in Fig.~\ref{fig:triangulated-surfaces}\,(c and f). These results demonstrate that differential geometry can model subterranean boundaries, handling topological changes (e.g., region merging) satisfactorily. The proposed techniques essentially transform irregularly-spaced points of  low resolution into continuously deformable surfaces that represent boundaries. Fig.~\ref{fig:bedded-surfaces} reveals the structure of the two geozones. The two vantage points show that $g_1$ lies above $g_2$, such bedded planes (layered geological formations) are common for iron ore deposits at the mine site.

\subsection{Predictive Value}
\begin{figure}[h]
\centering
\includegraphics[width=87mm]{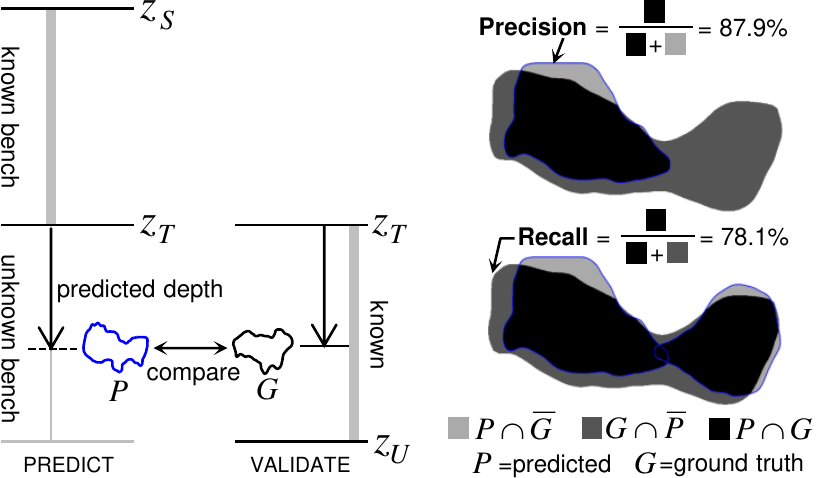}
\caption{Concept of precision and recall for spatial regions}
\label{fig:precision-recall-for-region}
\end{figure}

To assess the utility of slope information conveyed by the model, precision and recall rates are computed for predicted contours and compared against ground truth boundaries. The terrain extends from 690m down to 610m and is partitioned into disjoint 10m intervals (called benches) and processed top-down. Predictions are made using only information above the current bench floor. Therefore, predicted contours represent extrapolations from the known interval. For each comparison, data at the predicted depths from the bench below are withheld (not used to inform the prediction model) and these contours are used only for validation purpose at a given bench. Precision and recall rates are defined using set notations. The general setup is shown in Fig.~\ref{fig:precision-recall-for-region}. For predicted region $P_i$, precision is defined as $p(i)=\bigcup_{j\mid G_j\cap P_i}\left|P_i\cap G_j\right|/\left|P_i\right|$. For ground truth region $G_i$, recall is defined as $r(i)=\bigcup_{k\mid G_i\cap P_k}\left|G_i \cap P_k\right|/\left|G_i\right|$. The overall precision (resp., recall rate) at depth $d$ is the area-weighted average precision (resp., recall rate) of all contours at the specified depth. The depth varies from 0 to 10m in steps of 1.25m. Precision and recall statistics are computed under two conditions: (i) not using slope information [prediction employs zero-order hold], (ii) using the gradient field from the proposed model to obtain displacement estimates, i.e., combining the translation with the local deformation component estimated during spatial correspondence and contour metamorphosis, respectively.

\begin{figure}[h]
\centering
\includegraphics[width=76mm,trim={8mm 0mm 8mm 8mm},clip]{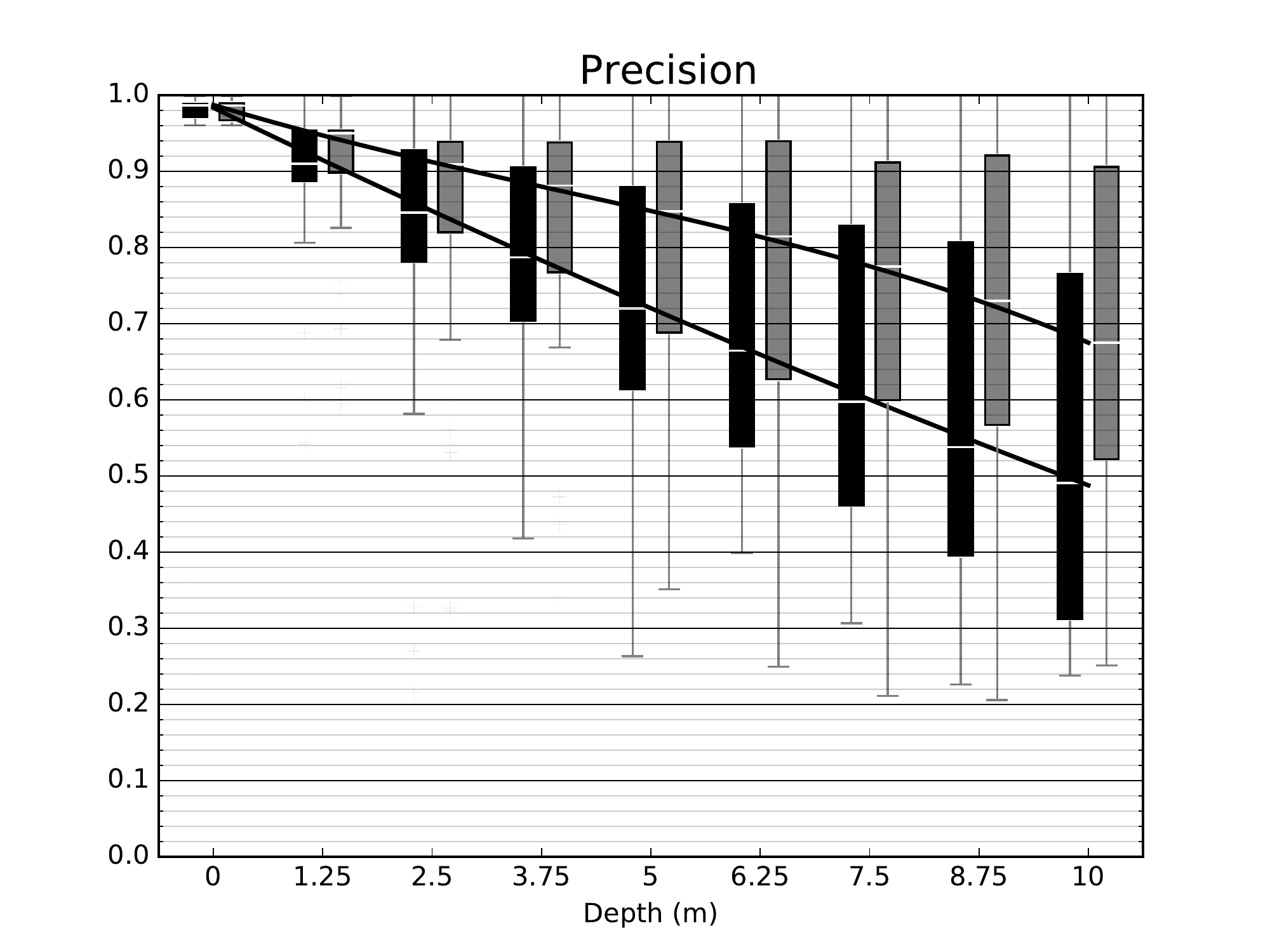}
\includegraphics[width=76mm,trim={8mm 4mm 8mm 8mm},clip]{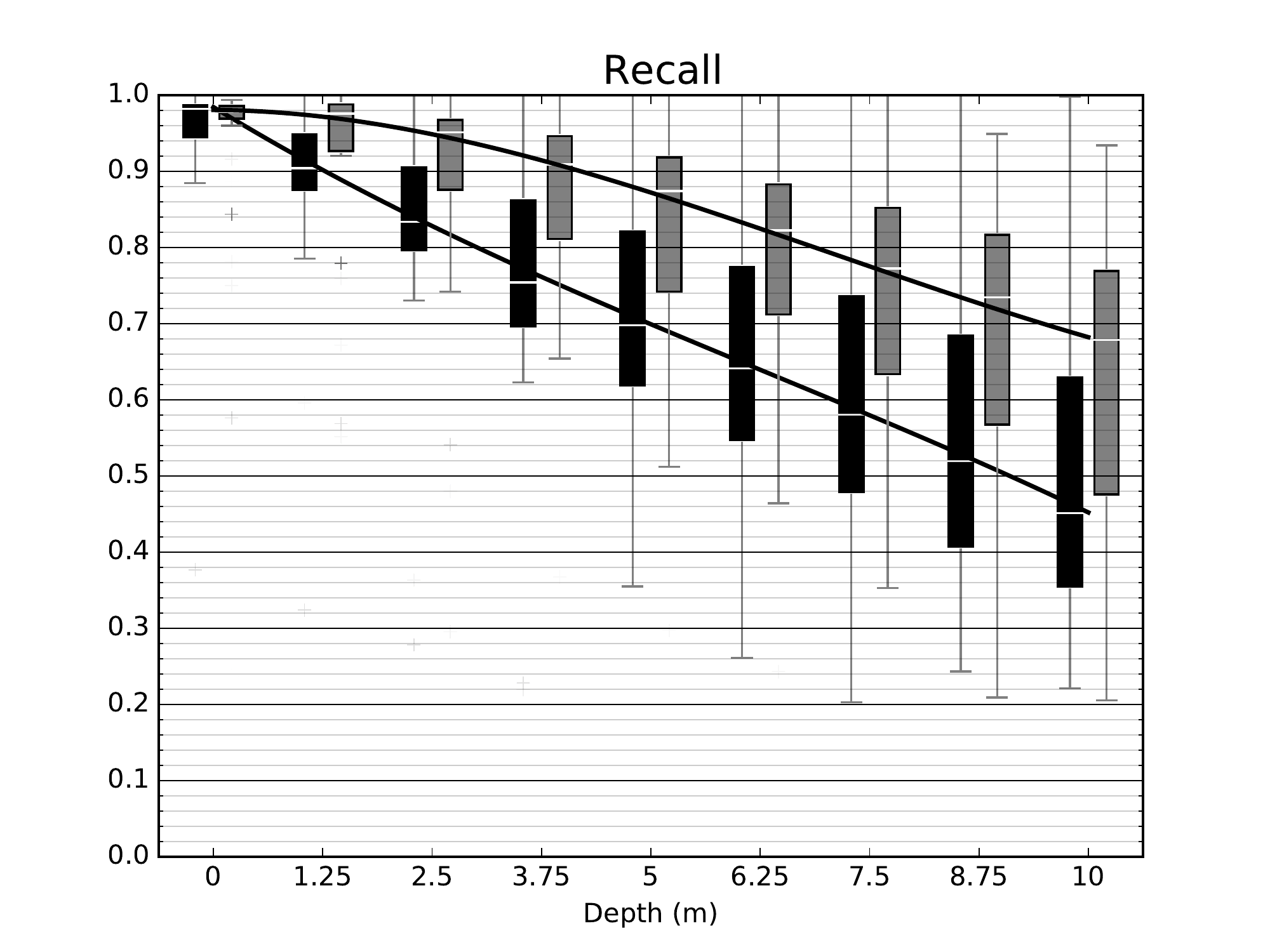}
\caption{Precision and recall rates for prediction: (gray) proposed model, (black) scenario where slope information is unavailable.}
\label{fig:model-precision-recall}
\end{figure}

Fig.~\ref{fig:model-precision-recall} demonstrates a consistent improvement in precision and recall rate when slope information is used in the model prediction. Significant gains in precision (12.7--19.2\%)  and recall (17.6--22.8\%) are observed at depths of 5 to 10 meter. These statistics are summarized in Table~\ref{tab:model-precision-recall}.

\begin{table}[h]
\center
\caption{Model precision and recall rates vs prediction depth}\label{tab:model-precision-recall}
\begin{tabular}{|c|ccc|ccc|}\hline
Depth (m) & \multicolumn{3}{c|}{Precision (\%)} & \multicolumn{3}{c|}{Recall (\%)}\\
& nil & model & gain & nil & model & gain\\ \hline
1.25 & 91.0 & \textbf{95.0} & +4.0 & 90.4 & \textbf{97.6} & +7.2\\ \hline
2.5 & 84.6 & \textbf{90.9} & +6.3 & 83.4 & \textbf{95.1} & +11.7\\ \hline
3.75 & 78.7 & \textbf{88.1} & +9.4 & 75.4 & \textbf{90.9} & +15.5\\ \hline
5.0 & 72.0 & \textbf{84.7} & +12.7 & 69.8 & \textbf{87.4} & +17.6\\ \hline
6.25 & 66.5 & \textbf{81.4} & +14.9 & 64.1 & \textbf{82.3} & +18.2\\ \hline
7.5 & 59.8 & \textbf{77.5} & +17.7 & 58.0 & \textbf{77.3} & +19.3\\ \hline
8.75 & 53.8 & \textbf{73.0} & +19.2 & 51.9 & \textbf{73.4} & +21.5\\ \hline
10.0 & 49.1 & \textbf{67.5} & +18.4 & 45.1 & \textbf{67.9} & +22.8\\ \hline
\end{tabular}
\end{table}

\section{Discussion}\label{sect:discuss}
These results demonstrate the feasibility of modeling subsurface boundary geometry using interdisciplinary techniques. In spite of the incomplete, sparse, causal and limited nature of the spatial patterns, computational physics (partial differential equations) and computer vision techniques can be combined to estimate directionality and produce a subsurface model that gives greater insight into the general structure of a geological domain. Modeling geological boundaries using differential geometry also offers other advantages. For ore reserve estimation, the volume of the triangulated surfaces may be computed easily from the vertices as outlined by Zhang \cite{zhang-01}. Given coherent slope estimates, directional predictions made by the model may be used in adaptive sampling \cite{ahsan-15} to make drilling more targeted and cost-effective. Indeed, various forms of analysis performed on those samples (including chemical assays and material classification by experts) can further constrain or correct inaccurate predictions, when geological boundaries need to be updated or learned further afield. This may be appreciated as a form of active learning in an exploration/exploitation framework. The surface slopes can also serve as a decision support tool for multi-agent scheduling and planning activities that involve drilling and excavation.

Several observations about the BGM system should be noted. First, although the boundary segmentation process has been designed to extract contours from sparse spatial patterns, it can take as input any sensible closed contour generated by humans or computers, such as lithological boundary detection using rock face image segmentation \cite{vasuki-17} and 2D decision boundaries from Gaussian process implicit surfaces \cite{williams-07} or SVM. Second, for spatial correspondence, it is possible to have humans in the loop to provide interactive estimates and perform component alignment in difficult cases. Third, the contour metamorphosis step may be preceded by geometric operations such as affine transform. For instance, a rotation between the source and target components can be estimated in the Fourier domain using the technique described by Nagashima \cite{nagashima-07}. Fourth, the directional estimates generated by the model can feedback into the system and act as a prior during region association. Finally, when data becomes more abundant, machine learning techniques can potentially be used to estimate certain parameters relating to surface dynamics (e.g., the anticipated lateral movement for certain regions).

\section{Conclusion}\label{sect:conclusion}
This paper presented a framework for modeling geological boundaries using differential geometry. The objective was to create subsurfaces from sparse spatial patterns and obtain coherent directional predictions along the boundary of the extracted surfaces. Under the model, the precision and recall rate for contour prediction improved on average by 15--20\%.  For boundary extraction, an edge map synthesis procedure was described. This converted sparse non-uniform data points to an image representation and facilitated the use of PDE-based techniques which operate on dense uniform 2D arrays. Gradient vector field and active contours were used to obtain regularized contours from unordered edge pixels. For spatial correspondence, region association and component alignment problems were examined, translation estimates were obtained using FFT cross-correlation under spatial constraints. Multi-source mult-target component relationships were explored using intersection graphs; strategies for obstacle avoidance were formulated from a resource allocation perspective. For contour metamorphosis, local surface deformation was modeled using PDE (described in Appendix~B). Surface slopes were estimated using normalized particle trajectories. The branching phenomenon was described: branching occurs during contour morphing when there is a significant mismatch in curvature between the source and target boundary segments. A curvelet backtracking algorithm was proposed to recover information lost during particle advection and thus overcome tracking failure caused by branching. In essence, the overall solution dealt with sparsity (irregularity), spatial constraints, and branching-induced tracking failure.

In a \textit{Nature} editorial, an opinion piece \cite{nature-interdisciplinary-15} characterized interdisciplinary research as a synthesis of different approaches into something unique. This captures the essence of this study which is to re-imagine how concepts from established areas can be synthesized and applied to a different field. This study had a narrow focus on estimating and exploiting the directionality of subsurface boundaries for more targeted drilling and exploration. Its uniqueness stems from applying differential geometry to sparse data. A different blend of technologies may be appropriate in a different setting where automation involves something different. Whilst research papers in engineering and the natural sciences have increasingly cited work outside their own disciplines \cite{vannoorden-15}, some areas remain unexplored. These areas can benefit from fresh perspectives and offer opportunities for meaningful collaboration \cite{nature-catalyse-15}. It is hoped this work can encourage more people to engage in and apply their expertise to various emerging and non-traditional fields.

\begin{appendix}

\section*{Appendixes}\label{sect:appendixes-brief}
Appendixes are included as \href{https://ieeexplore.ieee.org/ielx7/6287639/8600701/8891690/supplementalmaterial.pdf?tp=&arnumber=8891690}{Supplementary Material}. With open-access, this may be downloaded from \url{https://ieeexplore.ieee.org/ielx7/6287639/8600701/8891690/supplementalmaterial.pdf?tp=&arnumber=8891690}. The subject matters are outlined in Table~\ref{tab:appendixes-listing}.

\begin{table}[h]
\centering
\caption{List of appendixes (see \href{https://ieeexplore.ieee.org/ielx7/6287639/8600701/8891690/supplementalmaterial.pdf?tp=&arnumber=8891690}{\color{blue}Supplementary Material})}\label{tab:appendixes-listing}
\begin{tabular}{|c|l|}\hline
A & Theoretical treatment of active contours \& gradient vector field\\ \hline
B & Foundations for contour metamorphosis\\ \hline
\end{tabular}
\end{table}

\section*{Algorithms}
Algorithms are also included as \href{https://ieeexplore.ieee.org/ielx7/6287639/8600701/8891690/supplementalmaterial.pdf?tp=&arnumber=8891690}{Supplementary Material}. Table~\ref{tab:algorithms-listing} lists the algorithms and referenced sections.

\begin{table}[h]
\centering
\caption{List of algorithms (see supplemental material)}\label{tab:algorithms-listing}
\begin{tabular}{|c|l|c|}\hline
& Description & Ref.\,section\\ \hline
1 & Active contour evolution in GVF & Appendix \,A\\ \hline
2 & Edge map synthesis & \S\,\ref{sect:edgemap-synthesis}\\ \hline
3 & Signed distance function & \S\,\ref{sect:bgm-displacement-estimation-fft}\\ \hline
4 & Single-source single-target correspondence & \S\,\ref{sect:bgm-displacement-estimation-fft}\\ \hline
5 & Port allocation to handle spatial contention & \S\,\ref{sect:bgm-many-to-1-scenario}\\ \hline
6 & Multi-source single-target correspondence & \S\,\ref{sect:bgm-many-to-1-scenario}\\ \hline
7 & Single-source multi-target correspondence & \S\,\ref{sect:bgm-1-to-many-scenario}\\ \hline
8 & Multi-source multi-target correspondence & \S\,\ref{sect:bgm-many-to-many-scenario}\\ \hline
9 & Unified spatial correspondence strategy & \S\,\ref{sect:bgm-many-to-many-scenario}\\ \hline
10 & Wavefront curvelets backtracking & \S\,\ref{sect:curvelet-backtracking}, Fig.~\ref{fig:curvelets-illustrations}\\ \hline
11 & Wavefront curvelets branching analysis & \S\,\ref{sect:curvelet-backtracking}, Fig.~\ref{fig:curvelets-branching-analysis}\\ \hline
12 & Assembling particle trajectories (merging & \S\,\ref{sect:curvelet-backtracking}, Fig.~\ref{fig:curvelets-branching-analysis}\\
& existing and new trajectories from curvelets) & \\ \hline
\end{tabular}
\end{table}
\end{appendix}

\section*{Acknowledgment}
This work was supported by the Australian Centre for Field Robotics and the Rio Tinto Centre for Mine Automation. Professor Salah Sukkarieh and the reviewers are thanked for their suggestions and considered opinion which improved the quality of this manuscript.

\bibliographystyle{unsrt}  
\bibliography{ms}

\end{document}